\documentclass[12pt,nofootinbib]{article}
\pdfoutput=1
\usepackage{slashed}
\usepackage{amsmath,amssymb,graphicx,multicol} 
\usepackage{epsf}
\usepackage[nosort]{cite}
\usepackage{cancel}
\usepackage{framed}
\usepackage{multirow}
\usepackage{hyperref}
\usepackage{color}

\newcommand{\Tr}{\text{Tr}}

\newcommand{\beq}{\begin{eqnarray}}
\newcommand{\eeq}{\end{eqnarray}}

\newcommand{\centeron}[2]{{\setbox0=\hbox{#1}\setbox1=\hbox{#2}\ifdim

\wd1>\wd0\kern.5\wd1\kern-.5\wd0\fi \copy0

\kern-.5\wd0\kern-.5\wd1\copy1\ifdim\wd0>\wd1
                                       \kern.5\wd0\kern-.5\wd1\fi}}
\newcommand{\ltap}{\>\centeron{\raise.35ex\hbox{$<$}}
                               {\lower.65ex\hbox{$\sim$}}\>}
\newcommand{\gtap}{\>\centeron{\raise.35ex\hbox{$>$}}
                               {\lower.65ex\hbox{$\sim$}}\>}

\newcommand\ZZ{\hbox{\zfont Z\kern-.4emZ}}
\font\zfont = cmss10 

\begin{document}
\begin{titlepage}
\vspace*{-2.cm}
\begin{flushright}
{\small
CERN-PH-TH/2013-047 \\
KA-TP-06-2013 \\
PSI-PR-13-04
}
\end{flushright}
\vspace*{1.5cm}

\begin{center}
{\Large \bf  
Effective Lagrangian for a light Higgs-like scalar
}
\end{center}
\vskip0.5cm

\renewcommand{\thefootnote}{\fnsymbol{footnote}}

\begin{center}
{\large Roberto Contino$^{\,a}$, Margherita Ghezzi$^{\, a}$, Christophe Grojean$^{\, b,c}$, \\[0.5cm]
Margarete M\"uhlleitner$^{\, d}$ and Michael Spira$^{\, e}$}
\end{center}

\vskip 20pt

\begin{center}
\centerline{$^{a}$ {\small \it Dipartimento di Fisica,  Universit\`a di Roma La Sapienza and INFN, Roma, Italy}}
\vskip 5pt
\centerline{$^{b}$ {\small \it ICREA at IFAE, Universitat Aut\`onoma de Barcelona, E-08193 Bellaterra, Spain}}
\vskip 5pt
\centerline{$^{c}${\small \it Theory Division, Physics Department, CERN, Geneva, Switzerland}}
\vskip 5pt
\centerline{$^{d}${\small \it Institute for Theoretical Physics, Karlsruhe Institute of Technology, Karlsruhe, Germany}}
\vskip 5pt
\centerline{$^{e}${\small \it Paul Scherrer Institut, CH--5232 Villigen PSI, Switzerland}}
\end{center}

\vglue 1.0truecm

\begin{abstract}
\noindent 
We reconsider the effective Lagrangian that describes a light Higgs-like boson and better clarify a few 
issues which were not exhaustively addressed in the previous literature. In particular we highlight the strategy to
determine whether the dynamics responsible for the electroweak symmetry breaking is weakly or strongly interacting.
We also discuss how the effective 
Lagrangian can be implemented into automatic tools for the calculation of  Higgs decay rates and 
production cross sections. 
%
\end{abstract}

\end{titlepage}

\section{Introduction}

The exploration of the weak scale has marked an important step forward with the discovery by  the ATLAS~\cite{:2012gk} and CMS~\cite{:2012gu} 
collaborations of a boson with mass $m_h \simeq 125\,$GeV,  whose production cross section and decay rates are compatible with those predicted
for the Higgs boson of the Standard Model (SM).
At the same time, no hint of the existence of additional new particles has emerged yet, which might shed light on the origin of the electroweak symmetry 
breaking (EWSB).
One is thus faced with the problem of which is the best strategy to describe the properties and investigate the nature of the new boson $h$, beyond the
framework of the Standard Model.
In absence of a direct observation of new states,  our ignorance of the EWSB sector can be parametrized 
in terms of an effective Lagrangian for the light boson. Such an effective description is valid as long as 
 New Physics (NP) states appear at a scale $M \gg m_h$, and is based on an expansion in the number of fields and derivatives~\cite{weinberg}.
The detailed form of the effective Lagrangian depends on which assumptions are made. Considering that the observation made by the LHC experiments
is in remarkable agreement with the SM prediction, although within the current limited experimental precision, 
it is reasonable to assume that  $h$ is a CP-even scalar that  forms an $SU(2)_L$ doublet together with the longitudinal polarizations of the $W$ and $Z$, 
so that the   $SU(2)_L\times U(1)_Y$ electroweak symmetry is linearly realized at high energies.  
Under these assumptions the effective Lagrangian can be expanded into a sum of operators with increasing dimensionality, where the leading NP effects
are given by  dimension-6 operators.

The parametrization of the deviations of the Higgs couplings in terms of higher-dimension operators started more than two decades ago. The experimental 
observation of the Higgs boson, however, calls for a more detailed analysis. First, a compilation of a complete and updated list of constraints on the various 
Wilson coefficients is in need. Second, the rather precise estimation of the Higgs mass below the gauge boson thresholds necessitates a careful computation 
including off-shell effects that have not been incorporated up-to-now when the SM Lagrangian is supplemented by higher-dimensional operators. 
It is the purpose of this paper to perform such an updated analysis. We will also discuss in detail the implications of the custodial symmetry 
on the generalized Higgs couplings and clarify a few other issues which were not exhaustively addressed in the previous literature, like for example
the connection with the effective Lagrangian for a non-linearly realized electroweak symmetry.
Finally, a precise comparison of the Higgs couplings with the SM predictions can only be done when higher-order effects are included in a consistent way, 
and we will develop a strategy to this end.

The paper is structured as follows.
In Section~\ref{sec:effLag} we review the construction of the effective Lagrangian for a light Higgs doublet.
By means of a naive power counting we
estimate the coefficients of the various operators and review the most important bounds set on them
by present experimental results on electroweak (EW) and flavor observables. 
Focusing on Higgs physics, we then discuss in Section~\ref{sec:NPvsSM} the relative effect of the various operators on physical observables.
Such an analysis,  first proposed in Ref.~\cite{SILH}, will allow us to identify which operators can probe the  Higgs coupling strength
to the new states and which instead are sensitive only to the mass scale~$M$. This is of key importance to distinguish between weakly-coupled
UV completions of the Standard Model, like Supersymmetric (SUSY) theories, and theories where the EW symmetry is broken by a new strongly-interacting 
dynamics which forms the Higgs boson as a bound state~\cite{compositeHiggs,Contino:2003ve,Agashe:2004rs,SILH}.
These are the two most compelling scenarios put forward to solve the hierarchy problem
of the Standard Model.
We conclude the section by discussing how the assumption of a Higgs
doublet and linearly-realized $SU(2)_L \times U(1)_Y$ can be relaxed. We illustrate the non-linear effective Lagrangian valid for the 
case of a generic CP-even scalar $h$ and discuss the implications of custodial invariance. 
Section~\ref{sec:beyondtree} is devoted to clarify a few issues  related to the use of the effective Lagrangian beyond
the tree level. We present our concluding discussion in Section~\ref{sec:conclusions}. 
In the Appendices~\ref{app:SM}-\ref{app:CPodd} we collect useful formulas and 
give further details on the construction of the effective Lagrangian. 
The details of how we derived the bounds on the dimension-6 operators are reported in Appendix~\ref{app:EWfit}.

As an illustration of our analysis and to better demonstrate how
the effective Lagrangian can be implemented into automatic tools for the computation of physical quantities like Higgs 
production cross sections and decay rates, we have written {\tt eHDECAY}~\footnote{{\tt eHDECAY} is available at the following 
URL: \url{http://www-itp.particle.uni-karlsruhe.de/~maggie/eHDECAY/}}, a modified version of the program {\tt HDECAY}~\cite{hdecay},
which includes the full list of leading bosonic operators. We will describe the program in a separate companion paper~\cite{eHDECAYpaper}.


\section{Effective Lagrangian for a light Higgs doublet}
\label{sec:effLag}

The most general $SU(3)_C\times SU(2)_L\times U(1)_Y$-invariant Lagrangian for a weak doublet $H$ at the level of dimension-6 operators 
was first  classified in a systematic way in Refs.~\cite{Buchmuller:1985jz}. 
Subsequent analyses~\cite{others,Grojean:2006nn} pointed out the presence of some redundant operators, and a minimal and complete list of operators was finally
provided in Ref.~\cite{Grzadkowski:2010es}.
As recently discussed in Ref.~\cite{SILH},  a convenient basis of operators relevant for Higgs physics, assuming that the Higgs is a CP-even weak doublet (this assumption will be relaxed in Appendix~\ref{app:CPodd}) and the baryon and lepton numbers are conserved,  is the following:
\begin{equation}
\label{eq:effL}
{\cal L} = {\cal L}_{SM}  + \sum_i \bar c_i O_i \equiv {\cal L}_{SM} + \Delta {\cal L}_{SILH}  + \Delta {\cal L}_{F_1} +  \Delta {\cal L}_{F_2}  
\end{equation}
with
\begin{equation}
\label{eq:silh}
\vphantom{00}\hspace{-.7cm}
\begin{split}
\Delta {\cal L}_{SILH} =
\, & \frac{\bar c_H}{2v^2}\, \partial^\mu\!\left( H^\dagger H \right) \partial_\mu \!\left( H^\dagger H \right) 
+ \frac{\bar c_T}{2v^2}\left (H^\dagger {\overleftrightarrow { D^\mu}} H \right) \!\left(   H^\dagger{\overleftrightarrow D}_\mu H\right)  
- \frac{\bar c_6\, \lambda}{v^2}\left( H^\dagger H \right)^3 \\[0.2cm]
& + \left( \left( \frac{\bar c_u}{v^2}\,  y_{u}\, H^\dagger H\,   {\bar q}_L H^c u_R +  \frac{\bar c_d}{v^2}\,  y_{d}\, H^\dagger H\,   {\bar q}_L H d_R 
+ \frac{\bar c_l}{v^2}\,  y_{l}\, H^\dagger H\,   {\bar L}_L H l_R \right)  + {\it h.c.} \right)
\\[0.2cm]
& +\frac{i\bar c_W\, g}{2m_W^2}\left( H^\dagger  \sigma^i \overleftrightarrow {D^\mu} H \right )( D^\nu  W_{\mu \nu})^i
+\frac{i\bar c_B\, g'}{2m_W^2}\left( H^\dagger  \overleftrightarrow {D^\mu} H \right )( \partial^\nu  B_{\mu \nu})   
\\[0.2cm]
&
+\frac{i \bar c_{HW} \, g}{m_W^2}\, (D^\mu H)^\dagger \sigma^i (D^\nu H)W_{\mu \nu}^i
+\frac{i\bar c_{HB}\, g^\prime}{m_W^2}\, (D^\mu H)^\dagger (D^\nu H)B_{\mu \nu} 
\\[0.2cm]
&+\frac{\bar c_\gamma\,  {g'}^2}{m_W^2}\, H^\dagger H B_{\mu\nu}B^{\mu\nu}
   +\frac{\bar c_g \, g_S^2}{m_W^2}\, H^\dagger H G_{\mu\nu}^a G^{a\mu\nu}
\, , 
\end{split}
\end{equation}
\vspace{0.5cm}
\begin{equation}
\label{eq:silh2}
\vphantom{00}\hspace{-3.5cm}
\begin{split}
\Delta {\cal L}_{F_1} =
\, & \frac{i \bar c_{Hq}}{v^2}  \left(\bar q_L \gamma^\mu q_L\right)  \big( H^\dagger{\overleftrightarrow D}_\mu H\big)
+ \frac{i \bar c^\prime_{Hq}}{v^2}  \left(\bar q_L \gamma^\mu \sigma^i q_L\right)  \big(H^\dagger\sigma^i {\overleftrightarrow D}_\mu H\big) 
\\[0.2cm]
& + \frac{i \bar c_{Hu}}{v^2}  \left(\bar u_R \gamma^\mu u_R\right)  \big( H^\dagger{\overleftrightarrow D}_\mu H\big)
+ \frac{i \bar c_{Hd}}{v^2}  \left(\bar d_R \gamma^\mu d_R\right)  \big( H^\dagger{\overleftrightarrow D}_\mu H\big) \\[0.2cm]
& +\left(  \frac{i \bar c_{Hud}}{v^2}  \left(\bar u_R \gamma^\mu d_R\right)  \big( H^{c\, \dagger} {\overleftrightarrow D}_\mu H\big) +{\it h.c.} \right)
 \\[0.2cm]
& + \frac{i \bar c_{HL}}{v^2}  \left(\bar L_L \gamma^\mu L_L\right)  \big( H^\dagger{\overleftrightarrow D}_\mu H\big)
+ \frac{i \bar c^\prime_{HL}}{v^2}  \left(\bar L_L \gamma^\mu \sigma^i L_L\right)  \big(H^\dagger\sigma^i {\overleftrightarrow D}_\mu H\big)
 \\[0.2cm]
& + \frac{i \bar c_{Hl}}{v^2}  \left(\bar l_R \gamma^\mu l_R\right)  \big( H^\dagger{\overleftrightarrow D}_\mu H\big) \, ,
\end{split}
\end{equation}
\vspace{0.5cm}
\begin{equation}
\label{eq:silh3}
\begin{split}
\Delta {\cal L}_{F_2} =
\, & \frac{\bar c_{uB}\, g'}{m_W^2}\,  y_u \,  {\bar q}_L H^c \sigma^{\mu\nu} u_R \,  B_{\mu\nu}
 + \frac{\bar c_{uW}\, g}{m_W^2}\,  y_u \, {\bar q}_L  \sigma^i H^c \sigma^{\mu\nu} u_R  \, W_{\mu\nu}^i
 + \frac{\bar c_{uG}\, g_S}{m_W^2}\, y_u \,  {\bar q}_L H^c \sigma^{\mu\nu} \lambda^a u_R  \, G_{\mu\nu}^a 
\\[0.2cm]
\, &  +  \frac{\bar c_{dB}\, g'}{m_W^2}\, y_d \, {\bar q}_L H \sigma^{\mu\nu} d_R \, B_{\mu\nu}
+  \frac{\bar c_{dW}\, g}{m_W^2}\,   y_d \, {\bar q}_L \sigma^i H \sigma^{\mu\nu} d_R \, W_{\mu\nu}^i
+  \frac{\bar c_{dG}\, g_S}{m_W^2}\,  y_d \,  {\bar q}_L H \sigma^{\mu\nu}  \lambda^a d_R \, G_{\mu\nu}^a
\\[0.2cm]
\, & + \frac{\bar c_{lB}\, g'}{m_W^2}\,  y_l \,   {\bar L}_L H  \sigma^{\mu\nu}  l_R \, B_{\mu\nu}
+ \frac{\bar c_{lW}\, g}{m_W^2}\,  y_l \,  {\bar L}_L \sigma^i H  \sigma^{\mu\nu}  l_R \, W_{\mu\nu}^i
+ {\it h.c.}  
\end{split}
\end{equation}
The SM Lagrangian ${\cal L}_{SM}$ and our convention for the covariant derivatives and the gauge field strengths  are reported for completeness in 
Appendix~\ref{app:SM}. In particular, $\lambda$ is the Higgs quartic coupling and the weak scale at tree level is defined to be
\begin{equation}
\label{eq:vdef}
v\equiv\frac{1}{(\sqrt{2} G_F)^{1/2}} = 246\,\text{GeV}\, .
\end{equation}
By $i H^\dagger {\overleftrightarrow { D^\mu}} H $ we denote the Hermitian derivative 
$i H^\dagger  (D^\mu H) - i   (D^\mu H)^\dagger H$, and $\sigma^{\mu\nu} \equiv i [\gamma^\mu, \gamma^\nu]/2$.
The Yukawa couplings $y_{u,d,l}$ and the Wilson coefficients $\bar c_{i}$ in Eq.~(\ref{eq:silh2}) are matrices in flavor space, and a sum over flavors 
has been left understood. Note that the assumption of a CP-even Higgs implies that the coefficients $\bar c_u, \bar c_d$ and $\bar c_l$ are real. 
As specified in Eq.~(\ref{eq:effL}),  we will denote as $O_i$ the dimension-6 operator whose coefficient is proportional to $\bar c_i$.

Our higher-dimensional Lagrangian, which is supposed to capture the leading New Physics effects, counts 
12$\,(\Delta {\cal L}_{SILH})+ 8\,(\Delta {\cal L}_{F_1})+8\,(\Delta {\cal L}_{F_2}) = 28$ operators. Five extra bosonic operators,
\begin{equation}
\label{eq:other5bosonicops}
\begin{gathered}
\frac{\bar c_{3W}\,  g^3}{m_W^2}\, \epsilon^{ijk} W_{\mu}^{i\, \nu} W_{\nu}^{j\, \rho} W_{\rho}^{k\, \mu}\, , \qquad
\frac{\bar c_{3G}\,  g_S^3}{m_W^2}\, f^{abc} G_{\mu}^{a\, \nu} G_{\nu}^{b\, \rho} G_{\rho}^{c\, \mu} \, , \\[0.1cm]
\frac{\bar c_{2W}}{m_W^2} \left(D^\mu W_{\mu\nu}\right)^i \left(D_\rho W^{\rho\nu}\right)^i\, ,  \qquad
\frac{\bar c_{2B}}{m_W^2} \left(\partial^\mu B_{\mu\nu}\right) \left(\partial_\rho B^{\rho\nu}\right)\, , \qquad
\frac{\bar c_{2G}}{m_W^2} \left(D^\mu G_{\mu\nu}\right)^a \left(D_\rho G^{\rho\nu}\right)^a\, , 
\end{gathered}
\end{equation}
which affect the gauge-boson propagators and self-interactions  but with no effect on Higgs physics, should also be added 
to complete the operator basis, as well as 22 four-Fermi baryon-number-conserving operators.~\footnote{Notice that the last three operators
in Eq.~(\ref{eq:other5bosonicops}) can be rewritten in favor of three additional  independent four-Fermi operators, as in the basis
of Ref.~\cite{Grzadkowski:2010es}. The coefficients $\bar c_{2W}$, $\bar c_{2B}$ contribute respectively to the $W$ and $Y$ parameters 
defined in Ref.~\cite{Barbieri:2004qk}.}
A comparison with Ref.~\cite{Grzadkowski:2010es} shows that two 
of our operators are actually redundant. As we shall explain in more detail in Section~\ref{sec:NPvsSM} (see Eqs.~(\ref{eq:OWrewritten1}), (\ref{eq:OBrewritten})),  
it is well  known~\cite{Grojean:2006nn,Barbieri:1992dk} that two particular linear combinations of the fermionic operators in $\Delta {\cal L}_{F_1}$ are equivalent to 
pure  oblique corrections parametrized by the operators ${O}_T$, ${O}_W$ and ${O}_B$:
\begin{equation}
\label{eq:redferm}
{O}_{H\Psi}^Y \equiv \sum_{\psi} Y_\psi \, {O}_{H\psi}   \sim   O_T \, ,  O_B
\qquad  {\rm and} \qquad
{O}'_{Hq} + {O}'_{HL}  \sim  O_W  \, ,
\end{equation}
where the sum runs over all fermion representations, $\psi=q_L, u_R, d_R, L_L, l_R$, whose hypercharge has been denoted as $Y_\psi$.
These two linear combinations have then to be excluded from $ \Delta {\cal L}_{F_1}$, and we end up with exactly 53  linearly-independent operators as in 
Ref.~\cite{Grzadkowski:2010es}.~\footnote{For completeness we collect in Appendix~\ref{app:CPodd} also the extra 6 bosonic operators of dimension-six that 
are CP-odd.} Any other dimension-six operator can be obtained from these 53 operators by using the equations of motion, or equivalently by performing appropriate 
field redefinitions.~\footnote{In particular, the following identities hold:
\begin{equation}
\begin{split}
\frac{g^2}{4 m_W^2} \, H^\dagger H \, W_{\mu\nu}^i W^{i\, \mu\nu} & \equiv O_{WW} = O_W - O_B + O_{HB} - O_{HW}  + \frac{1}{4}\, O_\gamma \\[0.2cm]
\frac{g g^\prime}{4 m_W^2} \, H^\dagger \sigma^i H \, W_{\mu\nu}^i B^{\mu\nu} & \equiv O_{WB} = O_B - O_{HB} - \frac{1}{4}\, O_\gamma \, .
\end{split}
\end{equation}
\label{ftt:opidentity}
}

Even though our basis~(\ref{eq:silh})--(\ref{eq:silh3}) is equivalent to the one proposed in Ref.~\cite{Grzadkowski:2010es}, we advocate that it is more 
appropriate for Higgs physics for at least three reasons~\cite{SILH}: \textit{i)}~Generic models of New Physics generate a contribution to the 
oblique $\hat S$ parameter~\cite{PT,Barbieri:2004qk} 
at tree-level, which in the basis of Ref.~\cite{Grzadkowski:2010es}  would have to be encoded in the two fermionic operators ${O}_{H\psi}^Y$ and ${O}'_{Hq} + {O}'_{HL}$ 
even in the absence of direct couplings between the SM fermions and the New Physics sector. There is an advantage in describing the oblique corrections in terms of 
the operators in~(\ref{eq:silh}) rather than in terms of the operators with  fermionic currents, which generate vertex corrections and modify the Fermi constant.
\textit{ii)}~The basis~(\ref{eq:effL}) isolates the contributions to the decays $h \to \gamma \gamma$ (from ${O}_\gamma$)  and $h \to \gamma Z$ 
(from ${O}_\gamma$ and ${O}_{HW}-{O}_{HB}$) that occur only at the radiative level in minimally coupled theories. \textit{iii)}~Our basis of operators
is more appropriate to establish the nature of the Higgs boson and determine the strength of its interactions. For example, as we shall explain momentarily,
if the Higgs boson is a pseudo Nambu-Goldstone boson (pNGB) the coefficient of the operator $O_\gamma$,  hence the rate $h\to \gamma\gamma$,
is suppressed, while in the basis of Ref.~\cite{Grzadkowski:2010es} this
reflects into a cancellation in the linear combination  $4 \bar c_\gamma + (\bar c_{WW} - \bar c_{WB})$ (cf. footnote~\ref{ftt:opidentity}).

While a complete classification of the operators is essential, having a power counting to estimate their impact on physical
observables, hence their relative importance, is equally crucial.
In this sense a simple yet consequential observation was made in Ref.~\cite{SILH}: when expanding the effective Lagrangian in the number of fields
and derivatives,  any additional power of $H$ is suppressed by a factor $g_*/M \equiv 1/f$, where $g_* \leq 4\pi$ denotes the  coupling 
strength of the Higgs boson to New Physics states and $M$ is their overall mass scale;  any additional derivative instead costs a factor $1/M$. 
If the light Higgs boson is a composite state of the dynamics at the scale $M$, it is natural to expect $g_* \gg 1$, hence $f \ll M$,
which implies that operators with extra powers of $H$ give the leading corrections to low-energy observables. 
On the other hand, in weakly-coupled completions of the Standard Model where $g_* \sim g$, all operators with the same dimension
can be equally important.
A proper analysis of the experimental results through the language of  the effective Lagrangian can thus give indication on whether the
dynamics at the origin of electroweak symmetry breaking is weakly or strongly interacting.
According to the power counting of  Ref.~\cite{SILH},  one naively estimates ($\psi=u,d,l,q,L$)~\footnote{Notice that our normalization differs from the one of 
Ref.~\cite{SILH}, and it is more convenient than the latter for a model-independent implementation of Eq.~(\ref{eq:silh}) in a computer program.
The factor multiplying each operator in the effective Lagrangian has been conveniently defined such that the dependence on $M$ and  $g_*$ is 
fully encoded in the dimensionless coefficients $\bar c_i$.}
\begin{equation}
\label{eq:NDAestimates}
\begin{gathered}
\bar c_{H}, \bar c_{T}, \bar c_{6}, \bar c_{\psi} \sim O\!\left(\frac{v^2}{f^2}\right) , \quad 
\bar c_{W}, \bar c_{B} \sim O\!\left(\frac{m_W^2}{M^2} \right)\, , \quad
\bar c_{HW}, \bar c_{HB}, \bar c_{\gamma}, \bar c_{g} \sim  O\!\left(\frac{m_W^2}{16\pi^2 f^2}\right) \phantom{\, ,}
\\[0.5cm]
\bar c_{H\psi} , \bar c^\prime_{H\psi} \sim  O\!\left(\frac{\lambda_\psi^2}{g_*^2} \frac{v^2}{f^2}\right)\, ,
\quad \bar c_{Hud} \sim O\!\left(\frac{\lambda_u \lambda_d}{g_*^2} \frac{v^2}{f^2}\right)\, ,
\quad \bar c_{\psi W}, \bar c_{\psi B}, \bar c_{\psi G} \sim  O\!\left(\frac{m_W^2}{16\pi^2 f^2}\right)\, ,
\end{gathered}
\end{equation}
where  $\lambda_\psi$ denotes the coupling of a generic SM fermion $\psi$
to the new dynamics.
It should be stressed that these estimates are valid at the UV scale $M$, at which the effective Lagrangian is matched onto explicit models. 
Renormalization effects between $M$ and the EW scale mix  operators with the same quantum numbers,  and give in general
subdominant corrections to the coefficients.
We shall comment on these renormalization effects in Section~\ref{sec:beyondtree}.
Notice that the estimates of ${\bar c}_{W,B}$, ${\bar c}_{H\psi}$, ${\bar c}^\prime_{H\psi}$ and ${\bar c}_{T}$ apply when  these coefficients
are generated at tree-level.
However, specific  symmetry protections which might be at work in the UV theory, like for example $R$-parity in SUSY theories, 
can force the leading corrections to arise at the 1-loop level.

Equation~(\ref{eq:NDAestimates}) suggests that in the case of  a strongly-interacting light Higgs boson (SILH) the leading New Physics effects in Higgs 
observables are parametrized by the operators $O_{H,T,6,\psi}$, and,  if the SM fermions couple strongly
to the  new dynamics, by the fermionic operators of Eq.~(\ref{eq:silh2})~\cite{SILH}. 
Notice that, compared to the naive counting, $\bar c_{HW, HB, g, \gamma}$ are suppressed by an additional factor $(g_*^2/16\pi^2)$.
This is because the corresponding operators contribute to the coupling of on-shell photons and gluons to neutral particles and modify 
the gyromagnetic ratio of the $W$, and are thus generated only at the loop level in a minimally coupled theory.
Similarly, the dipole operators of Eq.~(\ref{eq:silh3}) are generated at the loop-level only, hence their estimates have an extra loop factor.

A special and phenomenologically motivated case is represented by theories where the Higgs doublet is a composite Nambu--Goldstone (NG) boson of a 
spontaneously-broken symmetry ${\cal G} \to {\cal H}$ of the strong dynamics~\cite{compositeHiggs,Contino:2003ve,Agashe:2004rs,SILH}.
For these models the scale $f$ must be identified with the decay constant associated with the spontaneous breaking, and the 
naive estimate of the Wilson coefficients $\bar c_i$ is modified by the request of  invariance under ${\cal G}$ in the limit of vanishing explicit breaking.
At the level of dimension-6 operators, $O_\gamma$, $O_g$, $O_6$, $O_{u,d,l}$ and the dipole operators of Eq.~(\ref{eq:silh3})
violate the shift symmetry $H^i\to H^i + \zeta^i$ ($\zeta^i=\, const.$) that is included as part of the ${\cal G}/{\cal H}$ transformations. 
This means that they cannot be generated in absence of an explicit breaking of the global symmetry.
It follows, in particular, that the naive estimates of the operators $O_\gamma$ and $O_g$ carry in this case
an additional suppression factor~\cite{SILH},
\begin{equation}
\label{eq:cgcga}
\bar c_{\gamma}, \bar c_g \sim O\!\left(\frac{m_W^2}{16\pi^2 f^2}\right) \!\times \frac{g_{\not G}^2}{g_*^2}\, ,
\end{equation}
where $g_{\not G}$ denotes any weak coupling that breaks the Goldstone symmetry (one of the SM weak couplings in minimal models, 
i.e. the SM gauge couplings or the Yukawa couplings).
The operators $O_6$, $O_\psi$, $O_{\psi G}$, $O_{\psi W}$, $O_{\psi B}$ have been defined so that their prefactor already includes
one spurion coupling, precisely  the Higgs quartic coupling $\lambda$ in $O_6$, and the Yukawa coupling $y_\psi$ in the other operators -- indeed,
both these couplings vanish for an exact NG boson. The estimates of the corresponding coefficients ${\bar c}_6$, ${\bar c}_{\psi}$, ${\bar c}_{\psi G}$, 
${\bar c}_{\psi W}$, ${\bar c}_{\psi B}$ are thus not modified.

In writing Eq.~(\ref{eq:silh}) we have  assumed that each of the operators $O_{u,d,l}$
is flavor-aligned with the corresponding fermion mass term, as required in order to avoid large Flavor-Changing Neutral Currents (FCNC) 
mediated by the tree-level exchange of the Higgs boson (see for example Ref.~\cite{Agashe:2009di} for a natural way to obtain this alignment). 
This implies one coefficient for the up-type quarks ($\bar c_u$), one for down-type quarks ($\bar c_d$), 
and one for the charged  leptons ($\bar c_l$), i.e. the ${\bar c}_{u,d,l}$ are  proportional to the identity matrix in flavor space.

\subsection{Current bounds on flavor-preserving operators}
\label{sec:bounds}

It is useful to review some of the most important constraints on the coefficients $\bar c_i$ that follow
from  current experimental results, such as electroweak precision tests, flavor data and low-energy precision measurements.
For simplicity, we  focus on the bounds on  flavor-conserving operators, keeping in mind that they can come also from
flavor-changing processes. 
For a  discussion of the bounds on flavor-violating operators see for example the recent review of Ref.~\cite{Isidori:2013ez}
as well as Ref.~\cite{Isidori:2010kg}.

Among the strongest bounds are those on operators that modify the vector-boson self-energies.
The operator $O_T$, for example, violates the custodial symmetry~\cite{Sikivie:1980hm} and contributes to the EW
parameter $\epsilon_1$~\cite{eps123}. 
From the EW fit performed in Ref.~\cite{Baak:2012kk}, it follows, with 95\% probability,
\begin{equation}
\label{eq:eps1}
\Delta \epsilon_1 \equiv \Delta\rho   = \bar c_T(m_Z) \, , \qquad  -1.5 \times 10^{-3}< \bar c_T(m_Z) < 2.2 \times 10^{-3} \, .
\end{equation}
Such a stringent bound
can be more naturally satisfied by assuming
that the dynamics at the scale $M$ possesses an (at least approximate) $SU(2)_V$ custodial invariance. In this case  $c_T(M)=0$, and a non-vanishing 
value will be generated through the renormalization-group (RG) flow of this Wilson coefficient down to $m_Z$ in the presence of
an explicit breaking of the custodial symmetry, as due for example to the Yukawa  or  hypercharge  couplings.
We will discuss these renormalization effects in more detail in Section~\ref{sec:beyondtree}.
Notice that all the other dimension-6 operators in the effective Lagrangian are (formally) custodially symmetric and their coefficients 
will not be suppressed at the scale~$M$.~\footnote{More precisely, for all the other operators the only violation of the 
custodial symmetry  comes from the explicit breaking due to the gauging of hypercharge. As such, this breaking is external to the EWSB dynamics, since it
comes from the weak gauging of its global symmetries. Formal invariance of the operators can be restored by
uplifting the hypercharge gauge field to a whole triplet of $SU(2)_R$. The top Yukawa coupling is another source of explicit custodial breaking.}
The electroweak precision tests also imply a strong bound on $O_W + O_B$~\cite{SILH}, since this linear combination contributes to the  parameter 
$\epsilon_3$~\cite{eps123}. With 95\% probability, one has~\cite{Baak:2012kk}:
\begin{equation}
\label{eq:eps3}
\Delta \epsilon_3 = \bar c_W(m_Z) + \bar c_B(m_Z) \, , \qquad 
  -1.4 \times 10^{-3} <  \bar c_W(m_Z) + \bar c_B(m_Z) < 1.9 \times 10^{-3} \, .
\end{equation}
From the tree-level estimate of ${\bar c}_{W,B}$ reported in Eq.~(\ref{eq:NDAestimates}), and assuming an approximate custodial invariance to suppress 
${\bar c}_T$ as explained above, it follows that Eqs.~(\ref{eq:eps1}) and (\ref{eq:eps3}) set a lower bound $M\gtrsim \;$a few TeV.
This bound is quite robust and can be avoided only in weakly-coupled UV completions where 
an extra symmetry protection suppresses the leading contribution to ${\bar c}_{W,B}$ 
by an additional loop factor. Notable examples are SUSY theories with $R$-parity.

The fermionic operators in Eq.~(\ref{eq:silh2}) are strongly constrained by $Z$-pole measurements, as they modify the couplings
of the $Z$ to quarks and leptons:
\begin{equation}
\label{eq:deltag}
\frac{\delta g_{L\psi}}{g_{L\psi}}= \frac{1}{2}\, \frac{\left(-\bar c_{H\Psi}  + 2 \, T_{3L}  \, \bar c'_{H\Psi}\right) }{T_{3L}  - Q \sin^2\!\theta_W} \, ,
\qquad
\frac{\delta g_{R\psi}}{g_{R\psi}}= \frac{1}{2}\, \frac{\bar c_{H\psi} }{Q \sin^2\!\theta_W}\, ,
\end{equation}
where $T_{3L}$ and $Q$ are respectively the $SU(2)_L$ and electric charges of the fermion $\psi$,
and $\Psi = \{ L, q\}$ is the $SU(2)_L$ doublet to which $\psi_L$ belongs.
We used the results of Ref.~\cite{Baak:2012kk} to perform a fit on the coefficients $\bar c_{H\psi}$, $\bar c_{H\Psi}$, $\bar c_{H\Psi}^\prime$. 
The details of our analysis can be found in Appendix~\ref{app:EWfit} (see also Ref.~\cite{Redi:2011zi}). 
In the case of light quarks ($u,d,s$) we find the following bounds
\begin{equation}
\label{eq:fermbounds1}
\begin{gathered}
-0.02 < \bar c_{Hq1} < 0.03 \, ,  \quad -0.002 < \bar c^\prime_{Hq1} < 0.003\, , \\[0.2cm]
-0.003 < \bar c_{Hq2} < 0.006\, ,  \quad -0.003 < \bar c^\prime_{Hq2} < 0.006\, ,  \\[0.2cm]
-0.008 < \bar c_{Hu} < 0.02 \, , \quad  -0.03 < \bar c_{Hd} < 0.02 \, , \quad  -0.03 < \bar c_{Hs} < 0.02 \, ,
\end{gathered}
\end{equation}
while a fit on leptons and heavy quarks ($c,b$) gives
\begin{equation}
\label{eq:fermbounds2}
\begin{gathered}
-0.0003 < \bar c_{HL}  + \bar c^\prime_{HL} < 0.002\, ,  \quad -0.002 < \bar c_{HL}  -\bar c^\prime_{HL}  < 0.004\, , \quad -0.0009 < \bar c_{Hl} < 0.001\, , \\[0.1cm]
-0.003 < \bar c_{Hq_2} -\bar c^\prime_{Hq_2} < 0.01\, , \quad  -0.01 < \bar c_{Hc} < 0.02 \, , \\[0.1cm]
-0.008 < \bar c_{Hq_3} +\bar c^\prime_{Hq_3} < 0.002\, , \quad -0.06 < \bar c_{Hb} < -0.009 \, .
\end{gathered}
\end{equation}
All the above bounds have 95\% probability and by the various coefficients we mean their values at the scale $m_Z$.
The weakest constraint is that on the operator $O_{Hb}$, which modifies the coupling of $b_R$ to the $Z$ boson.
The operator involving two right-handed top quarks, $O_{Ht}$, is  unconstrained by EW data,  but it is also not relevant
for the Higgs decays and will be neglected in the following. The coefficient ${\bar c}_{Htb}$ is severely constrained by the
$b\to s\gamma$ rate. 
Indeed, the expansion of $O_{Htb}$ around the vacuum contains a vertex of the type $Wt_Rb_R$, which at 1-loop
gives a chirally-enhanced contribution to the rate (see for example Ref.~\cite{Vignaroli:2012si}). 
We find, with 95\% probability: 
\begin{equation}
\label{eq:fermbounds3}
-0.4 \times 10^{-3} < \bar c_{Htb}(m_W) < 1.3 \times 10^{-3}\, .
\end{equation}
For a given $(v/f)$, the above bounds set a limit on the couplings of the SM fermions to the new dynamics, see Eq.~(\ref{eq:NDAestimates}). 
Unless the scale of New Physics is very large, or some specific symmetry protection is at work in the UV theory (see for example 
the discussion in Ref.~\cite{Redi:2011zi}), it follows that the SM fermions must be very weakly coupled to the new dynamics,
with the exception of the top quark.

The constraints on the dipole operators of Eq.~(\ref{eq:silh3}) come from the current experimental limits on electric dipole moments (EDMs) and 
anomalous magnetic moments.
The bounds on the neutron and mercury EDMs for example strongly constrain the dipole operators with $u$ and $d$ quarks.
By using the formulas of Ref.~\cite{Pospelov:2005pr} 
we find, with 95\% probability, that: 
\begin{equation}
\begin{gathered}
\label{eq:udEDM}
-7.01 \times 10^{-6} < \text{Im}({\bar c}_{uB}+{\bar c}_{uW}) < 7.86 \times 10^{-6} \, ,  \\[0.3cm]
-9.42 \times 10^{-7} < \text{Im}({\bar c}_{dB}-{\bar c}_{dW}) < 8.40 \times 10^{-7} \, , \\[0.3cm]
-1.62 \times 10^{-6} < \text{Im}({\bar c}_{uG})< 2.01 \times 10^{-6} \, ,  \\[0.3cm]
-7.71 \times 10^{-7}  < \text{Im}({\bar c}_{dG}) < 5.70 \times 10^{-7} \, ,
\end{gathered}
\end{equation}
where the coefficients are evaluated at the low-energy scale $\mu \sim 1\,$GeV.
According to the naive estimate (\ref{eq:NDAestimates}), 
for $O(1)$ CP-violating phases these results imply a bound on $(v/f)^2$ at the level of $10^{-3}$.
In natural extensions of the SM,  such a strong limit clearly points to the need of a symmetry protection mechanism.
For a discussion, see for example Ref.~\cite{Redi:2011zi} for the case of composite Higgs theories, and Ref.~\cite{Paradisi:2009ey} for the case of SUSY theories.

Among the heavier quarks the most interesting bounds are those on 
dipole operators with top quarks~\cite{Kamenik:2011dk}. These  come from the experimental limit on
the neutron EDM, 
\begin{equation}
\label{eq:topcEDM}
-1.39 \times 10^{-4} < \text{Im}({\bar c}_{tG}) < 1.21 \times 10^{-4} \, ,
\end{equation}
the $b\to s\gamma$ and $b\to sl^+l^-$ rates,
\begin{equation}
\label{eq:topEDM}
- 0.057 < \text{Re}({\bar c}_{tW}+{\bar c}_{tB}) - 2.65\, \text{Im}({\bar c}_{tW}+{\bar c}_{tB})  < 0.20 \, ,
\end{equation}
and the $t\bar t$ cross sections measured at the Tevatron and LHC,
\begin{equation}
\label{eq:ttbound}
-6.12 \times 10^{-3} < \text{Re}({\bar c}_{tG}) < 1.94 \times 10^{-3} \, . 
\end{equation}
All these bounds have 95\%  probability and have been derived by making use of the formulas reported in Ref.~\cite{Kamenik:2011dk}.~\footnote{The 
coefficients are evaluated at the following scales: 
$\mu =m_t$ (Eqs.~(\ref{eq:topcEDM}) and (\ref{eq:ttbound})),
$\mu = m_W$  (Eq.~(\ref{eq:topEDM})).
}
It is worth noting that the bounds of Eqs.~(\ref{eq:topEDM}) and (\ref{eq:ttbound}) are still about one order of magnitude weaker than 
the size of $\bar c_{tG}$, $\bar c_{tW}$ and $\bar c_{tB}$ expected from the naive estimate~(\ref{eq:NDAestimates}) with $(v/f)^2\sim 0.1$. 
Additional weaker constraints arise from the limits on anomalous top interactions based on  top decays and single top production. 
From the results of Ref.~\cite{AguilarSaavedra:2011ct} we find that, with 95\% probability:
\begin{equation}
\label{eq:anomtopcoupl}
-1.2 < \text{Re}( {\bar c}_{bW} ) < 1.1 \, , \qquad -0.01< \text{Re}( {\bar c}_{tW} ) < 0.02 \, .
\end{equation}
where the coefficients are evaluated at the scale $\mu \sim m_t$.

In the lepton sector, the current measurements and SM predictions of the muon~\cite{PDGreview,Davier:2010nc} and 
electron~\cite{Hanneke:2010au,Giudice:2012ms} anomalous magnetic moments 
and the limits on their EDMs~\cite{Bennett:2008dy,Hudson:2011zz}  imply the following 95\% probability bounds:
\begin{gather}
\begin{gathered}
\label{eq:RElepdipole}
-1.64 \times 10^{-2} < \text{Re}({\bar c}_{eB}-{\bar c}_{eW}) < 3.37 \times 10^{-3} \, ,  \\[0.2cm]
1.88 \times 10^{-4} < \text{Re}({\bar c}_{\mu B}-{\bar c}_{\mu W}) <  6.43 \times 10^{-4} \, , 
\end{gathered}
\\[0.4cm]
\begin{gathered}
\label{eq:IMlepdipole}
-2.97 \times 10^{-7}  < \text{Im}({\bar c}_{eB}-{\bar c}_{eW})<  4.51 \times 10^{-7} \, ,  \\[0.2cm]
-0.26 < \text{Im}({\bar c}_{\mu B}-{\bar c}_{\mu W})< 0.29 \, ,
\end{gathered}
\end{gather}
where the coefficients are evaluated at the relevant low-energy scale.
Notice that the non-vanishing value of $\text{Re}({\bar c}_{\mu B}-{\bar c}_{\mu W})$ follows from the known $\sim 3.5\sigma$ anomaly in the
$(g-2)$ of the muon (see Ref.~\cite{PDGreview} for an updated review).
Among the bounds of Eqs.~(\ref{eq:anomtopcoupl}), (\ref{eq:RElepdipole}), (\ref{eq:IMlepdipole}) only those on $\text{Im}(\bar c_{eB} - \bar c_{eW})$ 
and $\text{Re}(\bar c_{\mu B} - \bar c_{\mu W})$ have the sensitivity to probe the values naively expected for these coefficients as reported in 
Eq.~(\ref{eq:NDAestimates}). In particular, the first one sets an upper bound on $(v/f)^2$ of order $10^{-3}$ for an $O(1)$ CP phase.

\section{Estimates of the effects on physics observables}
\label{sec:NPvsSM}

 While the Lagrangian $\Delta {\cal L}=\Delta {\cal L}_{SILH}  + \Delta
{\cal L}_{F_1} +  \Delta {\cal L}_{F_2}$ is completely general, the basis of operators of Eqs.~(\ref{eq:silh})--(\ref{eq:silh3}) is particularly useful 
to characterize the interactions of the Higgs sector.
In fact, as already anticipated, one of the main results of Ref.~\cite{SILH} is that of identifying which operators, hence which observables, are sensitive to 
the strength of the Higgs  interactions, rather than merely to the value of the New Physics scale $M$.  
In what follows we will discuss this point in greater detail and, starting from the analysis of Refs.~\cite{SILH,princeton}, we will try to highlight 
a possible strategy to determine whether the dynamics behind the electroweak breaking is weak or strong. Our analysis will be based on the naive estimates 
of the Wilson coefficients at the matching scale. In the next Section, we will discuss how the running from the matching scale to the weak scale affects these estimates.

\subsection{Operators sensitive to a  strongly-interacting Higgs boson}

\enlargethispage{0.2cm}
Let us start by considering the effects of the operators $O_H$, $O_T$, $O_{u,d,l}$ and $O_6$: they
modify the tree-level couplings of the Higgs boson to fermions, vector bosons and to itself.
In the unitary gauge and upon canonical normalization of the Higgs kinetic term, the Lagrangian reads~\cite{Contino:2010mh}
\begin{equation}
\begin{split}
{\cal L} =
& \, \frac{1}{2} \partial_\mu h\, \partial^\mu h - \frac{1}{2} m_h^2 h^2 - c_3 \, \frac{1}{6} \left(\frac{3 m_h^2}{v}\right) h^3 + \dots
\\[0.3cm]
&  + m_W^2\,  W^+_\mu W^{-\, \mu} \left(1 + 2 c_W\, \frac{h}{v} + \dots \right)  + \frac{1}{2} m_Z^2\,  Z_\mu Z^\mu \left(1 + 2 c_Z\, \frac{h}{v} + \dots \right) 
\\[0.3cm]
& - \sum_{\psi = u,d,l} m_{\psi^{(i)}} \, \bar\psi^{(i)}\psi^{(i)} \left( 1 + c_\psi \frac{h}{v} + \dots \right) + \dots
\end{split}
\label{eq:effLagr}
\end{equation}
where the Higgs couplings $c_{i= W, Z, \psi, 3}$, have been defined such that $c_i = 1$ in the SM, and $v$~is defined by Eq.~(\ref{eq:vdef}). 
Their expressions  as functions of the coefficients of the effective Lagrangian~(\ref{eq:silh}) are given in Table~\ref{tab:coupvalues}.
%
\begin{table}
\vspace*{-1cm}
\begin{center}
\renewcommand{\arraystretch}{2.0}
\begin{tabular}{cccc}
\hline
Higgs couplings & $\Delta {\cal L}_{SILH}$ & MCHM4 & MCHM5 \\ \hline
$c_W$ & $1 -\bar c_H/2$ & $\sqrt{1-\xi}$ & $\sqrt{1-\xi}$ \\
$c_Z$ & $1 -\bar c_H/2 -  \bar c_T$ & $\sqrt{1-\xi}$ & $\sqrt{1-\xi}$ \\
$c_\psi \ \left(\psi=u,d,l\right)$ & $1- (\bar c_H/2 + \bar c_\psi)$ & $\sqrt{1-\xi}$ & $\displaystyle \frac{1-2\xi}{\sqrt{1-\xi}}$\\
$c_3$ & $1+ \bar c_6 - 3 \bar c_H/2$ & $\sqrt{1-\xi}$ & $\displaystyle \frac{1-2\xi}{\sqrt{1-\xi}}$ \\
$c_{gg}$ & $8\, (\alpha_s/\alpha_2) \,\bar c_{g}$ &  0 & 0 \\
$c_{\gamma\gamma}$ & $8\sin^2\!\theta_W \,\bar c_{\gamma}$& 0 & 0 \\
$c_{Z\gamma}$ & $\displaystyle \left(\bar c_{HB}-\bar c_{HW} - 8\, \bar c_\gamma \sin^2\!\theta_W\right) \tan\theta_W$ & 0 & 0 \\
$c_{WW}$ & $-2\,\bar c_{HW}$ & 0 & 0\\
$c_{ZZ}$ & $-2\left(\bar c_{HW} +  \bar c_{HB} \tan^2\!\theta_W - 4 \bar c_\gamma\tan^2\!\theta_W \sin^2\!\theta_W \right)$ 
& 0 & 0 \\
$c_{W\partial W}$ & $\displaystyle -2(\bar c_W+\bar c_{HW})$ & 0 & 0 \\
$c_{Z\partial Z}$ & $\displaystyle - 2(\bar c_W+ \bar c_{HW}) - 2\left(\bar c_B+\bar c_{HB}\right) \tan^2\!\theta_W$ & 0 & 0 \\
$c_{Z\partial \gamma}$ & $\displaystyle 2\left( \bar c_B+\bar c_{HB} - \bar c_W - \bar c_{HW}\right) \tan\theta_W$ & 0 & 0
\\[0.5cm]
\hline
\end{tabular}
\end{center}
\vspace{-.3cm}
\caption{\small 
The second column reports the values of the Higgs couplings $c_i$ defined in~Eq.~(\ref{eq:chiralL})  
in terms of the coefficients $\bar c_i$ of the effective Lagrangian $\Delta {\cal L}_{SILH}$.
The last two columns show the predictions of the MCHM4 and MCHM5 models in terms of $\xi = (v/f)^2$; 
the effects of the heavy resonances have been neglected for simplicity, so that only the couplings $c_{W,Z,\psi,3}$ are non-vanishing.
The auxiliary parameter $\alpha_2$ is defined by Eq.~(\ref{eq:alpha2}). Note that the previous version of this paper contains an erroneous factor 2 in the dependence of $c_Z$ on $\bar c_T$.
} 
\label{tab:coupvalues}
\end{table}
%
The shifts from the SM value are of order
\begin{equation}
\delta c_i \sim  \frac{g_*^2 v^2}{M^2}= \frac{v^2}{f^2} \, .
\end{equation}
Hence, measuring the Higgs couplings probes the strength of its interactions to the new dynamics. 
Notice that the effective description given by $\Delta {\cal L}$ neglects higher powers of ($H/f$), and is thus valid only if the
shifts in the Higgs couplings are small: $\delta c_i \sim (v/f)^2 \ll 1$.
If the Higgs doublet is the NG boson of a spontaneously broken symmetry ${\cal G} \to {\cal H}$, on the other hand,
it is possible to resum all  powers of ($H/f$) by making use of the invariance under (non-linear) ${\cal G}$ transformations. Such an improved
effective Lagrangian thus relies only on the expansion in the number of derivatives. 
For example, in models based on the   $SO(5)/SO(4)$ coset~\cite{Agashe:2004rs,Contino:2006qr} the couplings of the Higgs boson to 
$W$ and $Z$ are predicted to be 
$c_W = c_Z \equiv c_V = \sqrt{1-\xi}$, where $\xi \equiv (v/f)^2$. The couplings to fermions, on the other hand, are not uniquely fixed by the choice of the 
coset, but depend on how the SM fermions are coupled to the strong dynamics.
The last two columns of Table~\ref{tab:coupvalues}  report the predictions of the Minimal Composite Higgs Model MCHM4 \cite{Agashe:2004rs} 
and MCHM5 \cite{Contino:2006qr},  where the SM fermions couple linearly to composite operators transforming  as the spinorial and fundamental 
representations of $SO(5)$, respectively. For simplicity, the predictions are derived by including only the effects of the Higgs non-linearities, and
neglecting those from the heavy resonances, hence only the coefficients $c_V, c_\psi$ and $c_3$ are non-vanishing.
The models MCHM4 and MCHM5 will be considered as benchmarks in the rest of this work.

In general, a shift of the tree-level Higgs couplings of order $(v/f)^2$
implies that the theory gets strongly coupled at energies $\sim 4\pi f$, unless new weakly-coupled  physics states set in to regulate the energy growth of
the scattering amplitudes.
The dominant effect comes from the energy growth of the $V_LV_L \to V_L V_L$ ($V=W^\pm, Z^0$) scattering amplitudes, which become non-perturbative at the 
scale $\Lambda_s = 4\pi v/\sqrt{|\bar c_H|}$. A modified coupling
to the top quark leads instead to strong $V_LV_L\to t\bar t$ scattering at energies of order $\Lambda_s = 16\pi^2 v^2/(m_t \sqrt{|\bar c_u + \bar c_H|})$.
The scale of New Physics is thus  required to lie below, or at, such ultimate range of validity of the effective theory: $M \lesssim \Lambda_s$.

\subsection{Operators sensitive to the scale of New Physics}
\label{sec:opTR}

The operators $O_W$, $O_B$ can be generated at tree-level by the exchange of heavy particles, for example heavy spin-1 states.
In the unitary gauge they are written in terms of the following three operators~\footnote{Here and in the following, derivatives acting on operators
in the unitary gauge are covariant under local $U(1)_{em}$ transformations. Operators like $(\partial^\mu Z_{\mu\nu}) \gamma^\nu h$ or
$(\partial^\mu \gamma_{\mu\nu}) \gamma^\nu h$ obviously cannot be generated since they  break the $U(1)_{em}$ local symmetry.}
\begin{equation}
\label{eq:OWBunitary}
(D^\mu W^+_{\mu\nu}) W^{-\,\nu} h\, , \quad
(\partial^\mu Z_{\mu\nu}) Z^\nu h\, , \quad
(\partial^\mu \gamma_{\mu\nu}) Z^\nu h
\end{equation}
plus  terms with zero or two Higgs fields. 
The fact that there are three possible operators in the unitary gauge indicates that their coefficients are related 
by one identity  if the Higgs boson belongs to an $SU(2)$ doublet, see Eq.~(\ref{eq:identity2}). 
We will discuss this point in greater detail in Section~\ref{sec:custimpl}.

It is easy to see that $O_W$, $O_B$ give corrections to the tree-level Higgs couplings and  generate quartic interactions 
with one vector boson and two SM fermions that contribute to the three-body decays 
$h\to VV^{*} \to V \psi\bar\psi$.~\footnote{We thank Riccardo Rattazzi for pointing this out to us.}
Indeed, by making use of the equations of motion,~\footnote{For simplicity 
we have left a sum over all fermion representations $\psi$ understood in Eq.~(\ref{eq:eom}).}
\begin{equation}
\label{eq:eom}
i D^{\mu} W_{\mu\nu}^i  = g\, H^\dagger\frac{\sigma^i}{2} {\overleftrightarrow D}_\nu H - ig\, \bar \psi \frac{\sigma^i}{2} \gamma_\nu \psi\, ,
\qquad
i \partial^{\mu} B_{\mu\nu}  = \frac{g^\prime}{2}\, H^\dagger  {\overleftrightarrow D}_\nu H - ig^\prime\, \bar \psi Y \gamma_\nu \psi\, ,
\end{equation}
one can rewrite $O_W$ and $O_B$ as
\begin{align}
\label{eq:OWrewritten1}
O_W& = -2\,  O_H + \frac{4}{v^2} (H^\dagger H) |D_\mu H|^2 + O'_{Hq}+ O'_{HL} \\[0.1cm]
\label{eq:OBrewritten}
O_B & =  2 \tan^2\!\theta_W \left( -O_T + O^Y_{H\Psi} \right)  \, ,
\end{align}
where the linear combination $O^Y_{H\Psi}$ has been defined in Eq.~(\ref{eq:redferm}).
Upon the field redefinition $H \to H - 2 {\bar c}_W  (H^\dagger H) H/v^2$, the operator $ (H^\dagger H) |D_\mu H|^2$ can be rewritten in terms of those 
in Eq.~(\ref{eq:silh}).   Specifically, Eq.~(\ref{eq:OWrewritten1}) becomes:~\footnote{By means of Eqs.~(\ref{eq:OBrewritten}) and (\ref{eq:OWrewritten2}) it is 
thus always possible to remove $O_W$ and $O_B$ provided the coefficients of the other operators are shifted as follows: $\bar c_i \to \bar c_i + \Delta \bar c_i$, with
\begin{equation}
\begin{gathered}
\Delta \bar c_H = -6 \, \bar c_W , \quad \ 
\Delta \bar c_T = -2\tan^2\!\theta_W \, \bar c_B\, , \quad \ 
\Delta \bar c_6 = -8 \, \bar c_W , \quad \ 
\Delta \bar c_\psi =   2\, \bar c_W
\\[0.15cm]
\Delta \bar c_{Hq}^\prime =\Delta \bar c_{HL}^\prime = \bar c_W 
\\[0.15cm]
6\, \Delta \bar c_{Hq} = \frac{3}{2}\, \Delta \bar c_{Hu}  = -3\, \Delta \bar c_{Hd}  = -2\, \Delta \bar c_{HL}  
 = - \Delta \bar c_{Hl}  = - 2\tan^2\!\theta_W\, \bar c_B \, .
\end{gathered}
\end{equation}
}
\begin{equation}
\label{eq:OWrewritten2}
O_W = -6\, O_H + 2\, \left( (O_u + O_d + O_l) + h.c. \right) +- 8\, O_6 +  O'_{Hq}+ O'_{HL} \, .
\end{equation}
From the estimates of $\bar c_{W}, \bar c_B$ and $\bar c_H, \bar c_T, \bar c_\psi, \bar c_6$ in Eq.~(\ref{eq:NDAestimates}) one can see that the shifts 
to the tree-level  Higgs couplings  due to $O_W$, $O_B$ are of order $(m_W/M)^2$, hence
subdominant in the case of a strongly interacting Higgs boson. Notice that the couplings of the Higgs boson
to $W$ and $Z$ get different shifts from $O_B$  
(since $\Delta \bar c_T \not =0$). In practice, the constraint~(\ref{eq:eps3})
bounds this custodial-symmetry breaking effect down to an unobservable level, unless some fine tuning is in place in the combination $\bar c_W + \bar c_B$
so that $\bar c_B$ can be large.
Notice that despite the operator $O_T$ is generated after using the equations of motion, its contribution to $\Delta \epsilon_1$ (corresponding to 
a non-vanishing $\hat T$ parameter~\cite{PT,Barbieri:2004qk})  
is exactly canceled by the vertex correction implied by the linear combination of  fermionic operators which is also generated.~\footnote{See for example
Eq.~(9.10) of Ref.\cite{Barbieri:1992dk}.}
This is of course expected, since $O_{W}$, $O_B$ only contribute to $\epsilon_3$, and not to $\epsilon_1$.

In general, the contribution of $O_W$, $O_B$ to inclusive observables, in particular to the Higgs decay rates, is of order $(m_W^2/M^2)$:
\begin{equation}
\label{eq:inclusiveOWB}
\frac{\delta \Gamma(h\to VV)}{\Gamma(h\to VV)}\bigg|_{O_W,O_B} \sim O\!\left(\frac{m_W^2}{M^2}\right)\, ,
\end{equation}
where in this case $VV = W^{(*)}W^{*}, Z^{(*)}Z^{*}, Z^{(*)}\gamma, \gamma\gamma$.
This implies that these operators are sensitive only to the value of the scale of New Physics~$M$, and do not probe the coupling strength $g_*$.
From the quantitative side, the constraint~(\ref{eq:eps3}) suggests that their effects in inclusive Higgs decay rates is too small to be observable.
For example, we find that for small $\bar c_{W,B}$ the tree-level correction to the  $WW$ and $ZZ$ partial rates is well approximated 
by:~\footnote{Here and in the following our approximated formulas have been obtained by using \texttt{eHDECAY}~\cite{eHDECAYpaper}
with $m_h = 125\,$GeV.  QCD corrections to the decay rates  are fully included. Electroweak corrections are instead not included, since their effect on the 
numerical prefactor appearing in front of the coefficients $\bar c_i$ is of order  $(v^2/f^2) (\alpha_2/4\pi)$ and thus beyond the accuracy of our
computation. See Ref.~\cite{eHDECAYpaper}
for more details.}
\begin{equation}
\label{eq:DrateWB}
\frac{\Gamma(h\to W^{(*)}W^{*})}{\Gamma(h\to W^{(*)}W^{*})_{SM}} \simeq 1 + 2.2 \, \bar c_W \, , \qquad
\frac{\Gamma(h\to Z^{(*)}Z^{*})}{\Gamma(h\to Z^{(*)}Z^{*})_{SM}} \simeq 1 + 2.0 \left( \bar c_W + \tan^2\!\theta_W\, \bar c_B\right) \, . 
\end{equation}
Notice that despite its custodial invariance, the operator $O_W$ affects in a slightly different way the decay of the Higgs boson into $WW$ and $ZZ$, 
due to the fact that at least one of the two final vector bosons is off-shell.~\footnote{It is easy to check that for $m_h > 2 m_Z$ and on-shell decays 
one has:
\begin{equation}
\label{eq:DrateWBon}
\frac{\Gamma(h\to WW)}{\Gamma(h\to WW)_{SM}} \simeq 1 + 4 \, \bar c_W \, , \qquad
\frac{\Gamma(h\to ZZ)}{\Gamma(h\to ZZ)_{SM}} \simeq 1 + 4 \left( \bar c_W + \, \tan^2\!\theta_W\,  \bar c_B \right)\, . 
\end{equation}
These formulas coincide with those of Eqs.~(79)--(80) of Ref.~\cite{SILH}, which are thus valid only for on-shell decays.
}
At the one-loop level $O_W$ also contributes to the  Higgs decays into $Z\gamma$ and $\gamma\gamma$ (while $O_B$ does not). We find:
\begin{equation}
\frac{\Gamma(h\to Z\gamma)}{\Gamma(h\to Z\gamma)_{SM}} \simeq 1 + 4.2 \, \bar c_W \, , \qquad
\frac{\Gamma(h\to \gamma\gamma)}{\Gamma(h\to \gamma\gamma)_{SM}} \simeq 1 + 5.0 \, \bar c_W \, ,
\end{equation}
which agree with Eqs.~(82) and (83) of Ref.~\cite{SILH}.~\footnote{The easiest way to compute the one-loop contribution of $O_W$ to the $Z\gamma$ and 
$\gamma\gamma$ rates
is by using Eq.~(\ref{eq:OWrewritten1}) to rewrite this operator in terms of the others. Among the operators generated in this way, only $O_H$  gives a contribution.
Notice that if Eq.~(\ref{eq:OWrewritten2}) is used instead, one has to take into account also the contribution of $(O_u+O_d+O_l)$ and
the shift to the Fermi constant induced by $O'_{Hq}+ O'_{HL}$.
}
For $\bar c_{W,B} \sim 10^{-3}$ the above approximate formulas imply corrections too small to be observed at the LHC.
On the other hand, one could try to take advantage of the different predictions in terms of angular and invariant mass distributions which
are implied by the dimension-6 operators compared to the tree-level SM prediction. The most promising strategy could be in fact that
based on the analysis of the angular distributions of the final fermions~\cite{Choi:2002jk,DeRujula:2010ys,Bolognesi:2012mm}.
{In the ideal case in which one is able to kill completely the SM tree-level contribution  by means of appropriate kinematic cuts, 
the relative effect of NP becomes of order
\begin{equation}
\label{eq:DrateWBang}
\frac{d\Gamma(h\to VV)}{d\Omega}\Big/\left(\frac{d\Gamma(h\to VV)}{d\Omega}\right)_{SM} \lesssim 1 + \bar c_{W,B}\,   \frac{16\pi^2}{g^2}\, ,
\end{equation}
which might leave room for observable effects even for ${\bar c}_{W,B} \sim O(10^{-3})$.
Clearly, a more precise assessment of the efficiency of such a strategy  requires a dedicated analysis~\cite{WP}.

\subsection{Operators generated at the one-loop level}
\label{sec:op1L}

Let us now focus on the operators $O_{HW}, O_{HB}, O_\gamma$ and $O_{g}$, which  are generated at the one-loop level.
In the unitary gauge,  $O_{HW,HB,\gamma}$ are rewritten in terms of 
\begin{equation}
W^+_{\mu\nu} W^{-\, \mu\nu} h\, , \quad  Z_{\mu\nu} Z^{\mu\nu} h\, , \quad
\gamma_{\mu\nu} \gamma^{\mu\nu} h\, , \quad Z_{\mu\nu}\gamma^{\mu\nu}h
\end{equation}
plus other terms with zero or two Higgs fields. 
Since the coefficients of the above four operators are functions of $\bar c_{HW}$, $\bar c_{HB}$ and $\bar c_\gamma$, 
they are related  by one identity, see Eq.~(\ref{eq:identity1}). 
We will discuss this point in greater detail in Section~\ref{sec:custimpl}.

As implied from the naive estimates (\ref{eq:NDAestimates}), the contribution of $O_{HW,HB,\gamma}$ to the $WW$ and $ZZ$ inclusive rates
is of order ($VV=WW, ZZ$)
\begin{equation}
\frac{\delta\Gamma(h\to VV)}{\Gamma(h\to VV)}\bigg|_{O_\gamma,O_{HW},O_{HB}} 
\sim O\!\left(\frac{m_W^2}{16\pi^2 f^2}\right)\, .
\end{equation}
Although  such an effect  depends on the Higgs interaction strength, it is 
suppressed compared to Eq.~(\ref{eq:inclusiveOWB}) by a loop factor.
We find that the following approximate formulas hold~\footnote{
For $m_h > 2 m_Z$ and on-shell decays, we find instead
\begin{equation}
\label{eq:HWHBonshell}
\begin{split}
\frac{\Gamma(h\to WW)}{\Gamma(h\to WW)_{SM}}  & \simeq 1 + 8 \, \bar c_{HW}\, , \\[0.3cm]
\frac{\Gamma(h\to ZZ)}{\Gamma(h\to ZZ)_{SM}}  & \simeq 1 + 8 \, (\bar c_{HW} + \tan^2 \theta_W \, \bar c_{HB}) 
- 16 \tan^2 \theta_W \sin^2 \theta_W \, \bar c_\gamma\, .
\end{split}
\end{equation}
Comparing with the analog formulas in Eqs.~(79) and (80) of Ref.~\cite{SILH}, we find that in these latter there is a missing factor 2 and
the term proportional to $\bar c_\gamma$ was not included either. 
Notice also that the effect of the off-shellness of the gauge bosons is rather large, as one can  see by comparing
Eq.~(\ref{eq:HWHBonshell}) with Eq.~(\ref{eq:HWHBoffshell}).
}
\begin{equation}
\label{eq:HWHBoffshell}
\begin{split}
\frac{\Gamma(h\to W^{(*)}W^{*})}{\Gamma(h\to W^{(*)}W^{*})_{SM}}  & \simeq 1 + 3.7 \, \bar c_{HW}\, , \\[0.3cm]
\frac{\Gamma(h\to Z^{(*)}Z^{*})}{\Gamma(h\to Z^{(*)}Z^{*})_{SM}}  & \simeq 1 + 3.0 \left(\bar c_{HW} + \tan^2\!\theta_W\, \bar c_{HB}\right) 
- 0.26\, \bar c_\gamma\, .
\end{split}
\end{equation}
While the contribution due to $\bar c_{HB}$ and $\bar c_{\gamma}$ explicitly violates the custodial symmetry and thus
differentiates $WW$ from $ZZ$, the different numerical factor multiplying $\bar c_{HW}$ in the two formulas
above is due to the off-shellness of at least one of the two vector bosons, similarly to Eq.~(\ref{eq:DrateWB}).
Although there is currently no stringent bound on the coefficients $\bar c_{HW, HB,\gamma}$, the estimate~(\ref{eq:NDAestimates}) suggests that their correction 
to inclusive rates is unobservable at the LHC.  
As discussed in the previous section, on the other hand, a study of the angular and invariant mass distributions of these decays can
potentially uncover the effect of New Physics.  In particular, an
estimate similar to that of Eq.~(\ref{eq:DrateWBang}) can be  derived also for 
${\bar c}_{HW, HB, \gamma}$.

The processes  $h\to \gamma\gamma$, $h\to Z\gamma$ and $h\to gg$ (or equivalently $gg\to h$) 
can in principle test the Higgs interaction strengths much more powerfully, since they arise at the one-loop level in the SM.
Naively one expects:
\begin{equation}
\frac{\delta\Gamma(h\to gg,\gamma\gamma,Z\gamma)}{\Gamma(h\to gg,\gamma\gamma,Z\gamma)}\bigg|_{O_g,O_\gamma,O_{HW},O_{HB}} 
\sim O\!\left(\frac{v^2}{f^2}\right)\, .
\end{equation}
We find that the following approximate formulas hold to good accuracy for small ${\bar c}_i$'s:
\begin{equation}
\begin{split}
\frac{\Gamma(h\to gg)}{\Gamma(h\to gg)_{SM}} & \simeq 1 + 22.2 \, \bar c_g \, \frac{4\pi}{\alpha_{2}} 
\\[0.3cm]
\frac{\Gamma(h\to \gamma\gamma)}{\Gamma(h\to \gamma\gamma)_{SM}} & \simeq 1 - 0.54 \, \bar c_\gamma \, \frac{4\pi}{\alpha_{em}} \, , 
\\[0.3cm]
\frac{\Gamma(h\to Z\gamma)}{\Gamma(h\to Z\gamma)_{SM}} & \simeq 
1 + 0.19 \left( \bar c_{HW} - \bar c_{HB} + 8\, \bar c_\gamma \sin^2\!\theta_W \right) \frac{4\pi}{\sqrt{\alpha_2 \alpha_{em}}}\, , 
\end{split}
\end{equation}
where we have conveniently defined
\begin{equation}
\label{eq:alpha2}
\alpha_2 \equiv \frac{\sqrt{2}\, G_F m_W^2}{\pi}\, ,
\end{equation}
and by $\alpha_{em}$ we indicate the value of the running electromagnetic  coupling $\alpha_{em}(q^2=0)$ in the Thomson limit.
If the Higgs boson is a NG boson, the coefficients $\bar c_g$ and $\bar c_\gamma$ are further suppressed by a factor $(g_{\not G}/g_*)^2$, see Eq.~(\ref{eq:cgcga}),
where $g_{\not G}$ is a weak coupling.
This implies that in this class of theories
the corrections to $\Gamma(h\to \gamma\gamma)$ and $\Gamma(h \to gg)$
depend only on the scale of New Physics and not on the Higgs interaction strength. 
In fact, in the case of minimal models with linear couplings, like for example the MCHM4 and MCHM5,
the low energy theorem~\cite{Ellis:1975ap,Kniehl:1995tn}  implies that the 
leading contribution to the $\gamma\gamma$ and $gg$ decay rates from the virtual exchange of heavy fermions 
is additionally suppressed~\cite{Falkowski:2007hz,Low:2010mr,Azatov:2011qy,Gillioz:2012se} 
due to a cancellation between the effect parametrized by ${\bar c}_{g, \gamma}$ and the one
that follows from the shift in the top Yukawa coupling due to ${\bar c}_u$ and ${\bar c}_H$
(see Ref.~\cite{Azatov:2011qy} for  an interesting exception).
In general, in   theories with a pNGB Higgs boson the local corrections to the rates $\Gamma(h\to \gamma\gamma)$ and $\Gamma(h \to gg)$ 
from $O_\gamma$ and $O_g$ are expected  to be small and subdominant
compared to the effect from the modified tree-level Higgs couplings.

\subsection{Fermionic operators}

The fermionic operators in $\Delta {\cal L}_{F_1}$ are sensitive to the strength of the couplings of the Higgs boson and of the SM fermions  to the new dynamics.
They lead to contact corrections to the three-body decays $h\to VV^* \to V\psi\psi$ which are naively of order
\begin{equation}
\frac{\delta\Gamma(h\to V\bar \psi\psi)}{\Gamma(h\to V\bar \psi\psi)} \sim O\!\left(\frac{v^2}{f^2} \, \frac{\lambda_\psi^2}{g_*^2}\right)\, .
\end{equation}
Compared to the  corrections from $O_W$ and $O_B$, the effect of the fermionic operators is potentially  enhanced by a factor $(\lambda_\psi^2/g^2)$.
In practice, the possibility of large fermionic couplings $\lambda_\psi$ is strongly constrained by LEP, see Eqs.~(\ref{eq:fermbounds1})-(\ref{eq:fermbounds3}).
Scenarios in which a large degree of compositeness of  either the left- or right-handed quarks is not ruled out are generically those
in which the corresponding operators in $\Delta {\cal L}_{F_1}$ are not generated as due to some protecting symmetry 
(see for example Refs.~\cite{Redi:2011zi,Redi:2012uj,Barbieri:2012tu}). 
Large corrections to the inclusive rate of the three-body decays $h\to V\bar \psi\psi$ from $\Delta {\cal L}_{F_1}$ are thus excluded,
while the possibility of detecting the effects of these operators through the analysis of differential distributions should be explored,
similarly to what has been discussed for $O_{W}$ and $O_B$.

Among the dipole operators in $\Delta {\cal L}_{F_2}$, those with light fermions are already strongly constrained by current precision data,
but potentially sizable effects could still come from the operators involving the top quark.
For example, the contribution of $O_{tG}$ to $gg \to h$, $gg\to t\bar t$, $gg\to t\bar t h$ is of order $E^2/(16\pi^2 f^2)$, where $E$ is the
energy scale relevant in the process. More in detail
\begin{equation}
\label{eq:OtGestimates}
\frac{\delta\sigma (gg\to h)}{\sigma (gg\to h)} \sim {\hat c}_{tG}\, , \qquad
\frac{\delta\sigma (gg\to t\bar t)}{\sigma (gg\to t\bar t)} \sim {\hat c}_{tG}\, \frac{\sqrt{s}}{m_t}\, ,  \qquad
\frac{\delta\sigma (gg\to t\bar t h)}{\sigma (gg\to t\bar t h)} \sim {\hat c}_{tG} \, \frac{s}{m^2_t}\, , 
\end{equation}
where we have defined $\hat c_{tG} \equiv \text{Re}({\bar c}_{tG}) \, (m_t^2/m_W^2) \sim m_t^2/(16\pi^2 f^2) \simeq 3\times 10^{-3} (v^2/f^2)$.
Notice that the experimental limit on the  neutron EDM  puts an upper bound on the imaginary part of $\hat c_{tG}$ at the $10^{-4}$ level, 
see Eq.~(\ref{eq:topcEDM}), which indicates that this is currently the most sensitive experiment on $\text{Im}(\bar c_{tG})$.
Some mechanism is however required to suppress the imaginary parts of the dipole operators involving light fermions, in order to satisfy the
stringent constraints of Eq.~(\ref{eq:udEDM}). By the same mechanism also $\text{Im}(\bar c_{tG})$ could be suppressed, so that the 
processes of Eq.~(\ref{eq:OtGestimates}) are essential to probe the contribution of $O_{tG}$ due to $\text{Re}({\bar c}_{tG})$.
From Eq.~(\ref{eq:OtGestimates}) and the naive estimate of $\hat c_{tG}$ it follows that the most sensitive process is perhaps $gg\to t\bar t$, in particular
the events at large invariant mass, although a precision larger than the one currently achieved is required to constrain $(v/f)$.
To this aim,  the analysis of differential distributions and spin correlations could be a successful strategy~\cite{Degrande:2010kt,Kamenik:2011dk}. 
The NP contribution to the process $gg\to t\bar t h$ can in principle get the largest enhancement from a cut on $\sqrt{s}$, but the small
rate might limit the actual sensitivity achievable at the LHC~\cite{Degrande:2012gr}.
Finally, additional information comes from the experimental limits on top anomalous couplings obtained at the Tevatron and the LHC,
although their sensitivity on NP is expected to be much smaller by naive estimate.
The operator $O_{tW}$, in particular, gives the largest effect and generates the anomalous coupling 
$g_R (g/m_W) \bar b_L \sigma^{\mu\nu} W_{\mu\nu}^- t_R$~\cite{AguilarSaavedra:2011ct}.
Naively one expects $g_R = (4 m_t/m_W) \,\bar c_{tW} \sim m_t m_W/(16\pi^2 f^2) = 1.5 \times 10^{-3} \, (v/f)^2$, an effect too small to be observed
even for $f$ of order~$v$.

\subsection{Non-linear Lagrangian for a Higgs-like scalar}

Summarizing, by working in the unitary gauge and  in the basis of fermion mass eigenstates, 
the effective Lagrangian relevant for Higgs physics reads as follows~\cite{Contino:2010mh}
\begin{equation} 
\label{eq:chiralL}
\begin{split}
{\cal L} =
& \, \frac{1}{2} \partial_\mu h\ \partial^\mu h - \frac{1}{2} m_h^2 h^2 
- c_3 \, \frac{1}{6} \left(\frac{3 m_h^2}{v}\right) h^3 - \sum_{\psi = u,d,l} m_{\psi^{(i)}} \, \bar\psi^{(i)}\psi^{(i)} \left( 1 + c_\psi \frac{h}{v} + \dots \right)
\\[0.cm]
& + m_W^2\,  W^+_\mu W^{-\, \mu} \left(1 + 2 c_W\, \frac{h}{v} + \dots \right) + \frac{1}{2} m_Z^2\,  Z_\mu Z^\mu \left(1 + 2 c_Z\, \frac{h}{v} + \dots \right) + \dots
\\[0.3cm]
& + \left(  c_{WW}\,   W_{\mu\nu}^+ W^{-\mu\nu}  + \frac{c_{ZZ}}{2} \, Z_{\mu\nu}Z^{\mu\nu} + 
c_{Z\gamma} \, Z_{\mu\nu} \gamma^{\mu\nu}   + \frac{c_{\gamma\gamma}}{2}\, \gamma_{\mu\nu}\gamma^{\mu\nu} + \frac{c_{gg}}{2}\, G_{\mu\nu}^aG^{a\mu\nu} \right) \frac{h}{v}
\\[0.2cm]
& +  \Big( c_{W\partial W}\left(W^-_\nu D_\mu W^{+\mu\nu}+ h.c.\right)+c_{Z\partial Z}\,  Z_\nu\partial_\mu Z^{\mu\nu}
+  c_{Z\partial \gamma}\, Z_\nu\partial_\mu\gamma^{\mu\nu} \Big)\, \frac{h}{v} + \dots
\end{split}
\end{equation}
where, we recall, $v$ is defined in Eq.~(\ref{eq:vdef}).
We have shown only terms involving up to three bosonic fields, and we have omitted in particular
those involving fermions that follow from  $\Delta {\cal L}_{F_1}+\Delta {\cal L}_{F_2}$. Their form can be  easily derived from Eqs.~(\ref{eq:silh2}) 
and~(\ref{eq:silh3}).  The relations between the couplings appearing in Eq.~(\ref{eq:chiralL}) and the coefficients of the dimension-6 operators 
in Eq.~(\ref{eq:silh}) are reported in Table~\ref{tab:coupvalues}.
It is worth noting that the same Lagrangian (\ref{eq:chiralL}) applies also to the case in which the electroweak symmetry $SU(2)_L\times U(1)_Y$ is non-linearly 
realized and $h$ is a generic CP-even scalar, singlet of the custodial symmetry, not necessarily connected with the EW symmetry breaking.
Indeed, each of the terms in (\ref{eq:chiralL}), being invariant under local
$U(1)_{em}$ transformations, can be dressed up with the Nambu-Goldstone bosons that are eaten to form the longitudinal $W$ and $Z$ polarizations and made
manifestly  $SU(2)_L\times U(1)_Y$ gauge invariant~\cite{CCWZ} (see also Ref.~\cite{Burgess:1992gz}). 
The explicit expression in such a basis has been given in Refs.~\cite{Azatov:2012bz, Alonso:2012px} at the level of four-derivative operators. 
In this sense the effective Lagrangian (\ref{eq:chiralL}) is a generic tool to understand  the origin of the newly discovered boson and the role it plays in 
the electroweak symmetry breaking dynamics. 
It is valid for arbitrary values of the couplings~$c_i$ appearing in Eq.~(\ref{eq:chiralL}), and it can be used to make computations of observable quantities at a 
given order in an expansion in $E/M$ and in $\alpha_{SM}/4\pi$, where by the latter we indicate the generic SM loop expansion parameter. That is in full analogy 
with other well-known effective theories, see Ref.~\cite{weinberg}. It should be stressed that, according to a well established methodology and similarly
to Eq.~(\ref{eq:silh}), in this effective Lagrangian  all quantum fluctuations associated to short-length modes (high-energy modes) have already been considered 
and are parametrized by local operators with an increasing number of derivatives, while quantum fluctuations (loop diagrams) involving the light modes 
still have to be taken into account.  For instance, top loops will give an additional contribution to the on-shell $h$-gluon-gluon coupling.
While Eq.~(\ref{eq:chiralL}) is general, the effective Lagrangian (\ref{eq:silh}) assumes that $h$ is part of an $SU(2)_L$ doublet and further relies on the expansion 
in powers of $H/f$. As such, it is  valid only in the limit of small deviations of the Higgs couplings from their SM values and up to
corrections of order $O(v^2/f^2)$.  

\subsection{Implications of custodial symmetry}
\label{sec:custimpl}

Another difference between the non-linear Lagrangian~(\ref{eq:chiralL}) and the  SILH Lagrangian~(\ref{eq:silh}) is that the first one
contains two more free parameters. This means that there are two relations  among the couplings of Eq.~(\ref{eq:chiralL})
which hold at the level of dimension-6 operators if the Higgs is part of a doublet. 
As noticed in Sections~(\ref{sec:opTR}) and (\ref{sec:op1L}), the first identity relates $c_{WW}$, $c_{ZZ}$,
$c_{Z\gamma}$ and $c_{\gamma\gamma}$, while the second relates $c_{W\partial W}$, $c_{Z\partial Z}$ and $c_{Z\partial \gamma}$. They read:
\begin{align}
\label{eq:identity1}
c_{WW}  - c_{ZZ} \cos^2\!\theta_W & = c_{Z\gamma} \sin 2\theta_W + c_{\gamma\gamma}  \sin^2\!\theta_W \\[0.4cm]
\label{eq:identity2}
c_{W\partial W}  - c_{Z\partial Z} \cos^2\!\theta_W & = \frac{c_{Z\partial \gamma}}{2} \sin 2\theta_W \, .
\end{align}
In fact both identities are not special to the case in which the Higgs is a doublet, but are a general consequence of custodial symmetry.
This latter is accidental in the SILH Lagrangian if one restricts to the operators that lead to derivative couplings of the Higgs to vector bosons. Starting at the dimension-8 order, it is possible to write custodial-breaking operators that lead to couplings that violate the relations~(\ref{eq:identity1}) and~(\ref{eq:identity2}). 
For instance
\begin{equation}
	\label{eq:dim8}
\frac{\bar c_{8WW} \, g^2}{m_W^2 v^2} \left( H^\dagger W_{\mu\nu}^a \sigma^a H \right) \left( H^\dagger W^{b\, \mu\nu} \sigma^b H \right)
+
\frac{i \bar c_{8W} \, g}{v^2 m_W^2 }\left( H^\dagger \sigma^a H \right) ( D^\mu  W_{\mu \nu})^a  \left( H^\dagger \overleftrightarrow {D^\nu} H \right ) 
\end{equation}
gives rise to
\begin{equation}
\begin{gathered}
c_{Z\partial Z}  = - 4 \bar c_{8W}, \qquad
c_{Z\partial \gamma}  = - 4 \tan \theta_W \bar c_{8W} \, , \\[0.2cm]
c_{ZZ}  = 8 \cos^2 \!\theta_W \bar c_{8WW} \, , \qquad
c_{Z\gamma}  = 4 \sin 2 \theta_W \bar c_{8WW}\, , \qquad
c_{\gamma \gamma} = 8 \sin^2 \!\theta_W \bar c_{8WW} \, , 
\end{gathered}
\end{equation}
and the relations (\ref{eq:identity1}) and~(\ref{eq:identity2}) are not fulfilled.~\footnote{The two operators in~(\ref{eq:dim8}) give rise to the oblique parameter $\hat U$, see for instance Ref.~\cite{Barbieri:2004qk}: $\hat U=-\bar c_{8W}-2\bar c_{8HW}$ while $\hat S=\bar c_{8HW}$.}

A third relation holds on the non-derivative couplings $c_W$ and $c_Z$ if one assumes that custodial symmetry is an invariance of the 
Lagrangian~(\ref{eq:silh}), so that $\bar c_T =0$; it reads:
\begin{equation}
\label{eq:identity3}
c_W =c_Z \, .
\end{equation}
As said above, while all three identities (\ref{eq:identity1}), (\ref{eq:identity2}) and (\ref{eq:identity3}) are a consequence of custodial symmetry, 
the first two are accidental at the level of dimension-6 operators if the Higgs is part of a doublet.

To show that Eqs.~(\ref{eq:identity1}), (\ref{eq:identity2}) and (\ref{eq:identity3}) follow from custodial invariance,
let us consider the case in which the EWSB dynamics has a global $SU(2)_L \times SU(2)_R$ symmetry, and imagine to fully gauge this group
by enlarging the hypercharge to a whole triplet of $SU(2)_R$. In this case the diagonal custodial $SU(2)_V$ is exact even though $g^\prime \not = g$. 
The left and right gauge fields couple to the conserved currents of $SU(2)_L \times SU(2)_R$ and the  interactions among two gauge fields
and the Higgs boson are fully characterized  in momentum space by three form factors:
\begin{equation}
\label{eq:LRff}
(\Gamma_{LL})^{\mu\nu}_{ij}(p_1,p_2) \, L_\mu^{i} L_\nu^j  h +(\Gamma_{LR})^{\mu\nu}_{ij}(p_1,p_2) \, L_\mu^{i} R_\nu^j h 
+ (\Gamma_{RR})^{\mu\nu}_{ij}(p_1,p_2) \, R_\mu^{i} R_\nu^j h \, .
\end{equation}
Here $p_1$, $p_2$ are the momenta of the gauge fields and each form factor 
can be computed in terms of a Green function with two conserved currents, $\Gamma^{\mu\nu}_{ik} = \langle J^\mu_i J^\nu_k | h \rangle$.
In addition to the usual massive $W$ and $Z$ bosons, which form a triplet $\hat V^i_\mu$ of the custodial group, in this case there is a whole
triplet of massless $SU(2)_V$ gauge fields (the photon plus its charged companion), $V^i_\mu$.
The mass eigenstates $V_\mu$ and $\hat V_\mu$ are related to the left and right gauge fields through a rotation by an angle $\theta_W$,
where $\tan\theta_W = g^\prime/g$. Their cubic interactions with the Higgs boson are thus characterized by three form factors,
which are  linear combinations of those in Eq.~(\ref{eq:LRff}):
%
%
\begin{equation}
\label{eq:VVhatff}
\begin{split}
\Gamma_{VV}& = \sin^2\!\theta_W \,\Gamma_{LL} + \frac{\sin 2\theta_W}{2} \left(\Gamma_{LR}+\Gamma_{RL}\right) +  \cos^2\!\theta_W \,\Gamma_{RR} \\
\Gamma_{\hat VV} & = \frac{\sin 2\theta_W}{2}\Gamma_{LL}+ \left( \cos^2\!\theta \, \Gamma_{LR} - \sin^2\!\theta \, \Gamma_{RL}\right)   - \frac{\sin 2\theta_W}{2}\Gamma_{RR} \\
\Gamma_{\hat V \hat V} & = \cos^2\!\theta_W \,\Gamma_{LL}  - \frac{\sin 2\theta_W\,}{2}\left(\Gamma_{LR}+\Gamma_{RL}\right)+  \sin^2\!\theta_W \,\Gamma_{RR} \, ,
\end{split}
\end{equation}
where we have defined $\Gamma_{RL}^{\mu\nu}(p_1,p_2) \equiv \Gamma_{LR}^{\nu\mu}(p_2,p_1)$.
Notice, in particular, that in this case the same form factor $\Gamma_{\hat V\hat V}$ describes the interaction of two $W$'s and two $Z$'s to the Higgs boson, 
as due to custodial invariance. 

The physical limit where only $SU(2)_L \times U(1)_Y$  is gauged is obtained by simply switching off the unphysical $R^{1,2}_\mu$ fields. 
The interactions of two neutral vector bosons to the Higgs are still described by the relations of Eq.~(\ref{eq:VVhatff}), where 
$\Gamma_{ZZ} = \Gamma_{\hat V \hat V}$, $\Gamma_{\gamma\gamma} = \Gamma_{VV}$ and $\Gamma_{Z\gamma} = \Gamma_{\hat V V}$.
In the charged sector, instead, the $W$ corresponds to a pure left gauge field, since it has no mixing with right-handed ones. This implies that
its form factor is given by the last formula of  Eq.~(\ref{eq:VVhatff}) with $\theta_W = 0$, that is: $\Gamma_{WW} = \Gamma_{LL}$.
The four physical form factors are linear combinations of  the three defined in Eq.~(\ref{eq:LRff}), and are thus related by one identity:
\begin{equation}
\label{eq:custidentity}
\Gamma^{\mu\nu}_{WW}(p_1,p_2) - \Gamma^{\mu\nu}_{ZZ}(p_1,p_2) \cos^2\!\theta_W = 
\left( \Gamma^{\mu\nu}_{Z\gamma}(p_1,p_2) + \Gamma^{\nu\mu}_{Z\gamma}(p_2,p_1) \right) \frac{\sin 2\theta_W}{2} 
+ \Gamma^{\mu\nu}_{\gamma\gamma}(p_1,p_2) \sin^2\!\theta_W \, .
\end{equation}
Notice that this relation is a consequence of our initial assumption of $SU(2)_L \times SU(2)_R$ invariance of the EWSB dynamics.
The custodial $SU(2)_V$ is broken in this case only by the gauging of hypercharge. For $g^\prime = 0$ the custodial symmetry is unbroken
and Eq.~(\ref{eq:custidentity}) implies $\Gamma_{WW} = \Gamma_{ZZ}$.
It is straightforward to derive the relations (\ref{eq:identity1}), (\ref{eq:identity2}) and (\ref{eq:identity3}) from Eq.~(\ref{eq:custidentity}). 
At quadratic order in the momenta, the form factors can be computed from the effective Lagrangian~(\ref{eq:chiralL}); one has:
\begin{equation}
\label{eq:ffdecomposition}
\begin{split}
\Gamma_{WW}^{\mu\nu}(p_1,p_2) = \, & 2 m_W^2 c_W \, \eta^{\mu\nu} - 2 c_{WW} P_{12}^{\mu\nu} - c_{W\partial W} \left( P_1^{\mu\nu} + P_2^{\mu\nu} \right)
\\[0.2cm]
\Gamma_{ZZ}^{\mu\nu}(p_1,p_2) = \, & 2 m_Z^2 c_Z \, \eta^{\mu\nu} - 2 c_{ZZ} P_{12}^{\mu\nu} - c_{Z\partial Z} \left( P_1^{\mu\nu} + P_2^{\mu\nu} \right)
\\[0.2cm]
\Gamma_{Z\gamma}^{\mu\nu}(p_1,p_2) = \, & - 2 c_{Z\gamma} P_{12}^{\mu\nu} - c_{Z\partial \gamma} P_2^{\mu\nu} 
\\[0.2cm]
\Gamma_{\gamma\gamma}^{\mu\nu}(p_1,p_2) = \, & - 2 c_{\gamma\gamma} P_{12}^{\mu\nu}\, ,
\end{split}
\end{equation}
where we have defined  $P_1^{\mu\nu} \equiv \eta^{\mu\nu} p_1^2 - p_1^\mu p_1^\nu$, $P_2^{\mu\nu} \equiv \eta^{\mu\nu} p_2^2 - p_2^\mu p_2^\nu$
and $P_{12}^{\mu\nu} \equiv \eta^{\mu\nu} p_1\!\cdot\! p_2 - p_1^\nu p_2^\mu$. 
This is in fact the most general decomposition  which follows at the $O(p^2)$ level for an on-shell Higgs boson
by assuming CP invariance and requiring that: \textit{i)} the $\Gamma_{WW}$, $\Gamma_{ZZ}$ and $\Gamma_{\gamma\gamma}$ form factors 
are symmetric under the exchange $\{ p_1, \mu \} \leftrightarrow \{ p_2 , \nu\}$; \textit{ii)} the $\Gamma_{\gamma\gamma}$ and
$\Gamma_{Z\gamma}$ form factors satisfy the Ward identities implied by $U(1)_{em}$ local invariance:
\begin{equation}
p_{1 \mu} \Gamma^{\mu\nu}_{\gamma\gamma}(p_1,p_2) = 0 = p_{2\nu} \Gamma^{\mu\nu}_{\gamma\gamma}(p_1,p_2) \, ,
\qquad 
p_{2\nu} \Gamma^{\mu\nu}_{Z\gamma}(p_1,p_2) = 0\, .
\end{equation}
Additional structures proportional to $p_{1\mu}$ and $p_{2\nu}$ can be omitted since they give vanishing contributions 
both when the vector bosons are on-shell and when they decay into a pair of fermions by coupling to the corresponding conserved current.
Inserting Eq.~(\ref{eq:ffdecomposition}) into (\ref{eq:custidentity}) one then obtains the identities~(\ref{eq:identity1}), (\ref{eq:identity2}) and (\ref{eq:identity3}).

From the above discussion it follows that if  custodial symmetry is an invariance of the EWSB dynamics, the effective Lagrangians~(\ref{eq:chiralL})
and (\ref{eq:silh}) have the same number of free parameters, in terms of which all observables can be computed.
This is true also if one considers the fermionic operators (for a Higgs doublet these are listed in Eqs.~(\ref{eq:silh2}) and (\ref{eq:silh3})), 
as long as one focuses on terms with one Higgs boson.
This means that by using  single-Higgs processes alone, one cannot distinguish the case in which the Higgs boson is part of a doublet from the more
general situation. The only possible strategy to this aim is exploiting the connection among processes with zero, one and two
Higgs bosons which is implied by the Lagrangian~(\ref{eq:effL}) at $O(v^2/f^2)$ and does not hold in the case of the more general non-linear Lagrangian.
As a consequence of such connection,
the bounds that EW and flavor data set on operators with zero Higgs fields severely constrain the size of the NP effects 
in Higgs processes, as discussed in Section~(\ref{sec:bounds}). If one were to find that single-Higgs processes violate these constraints, this would be
an indication that the Higgs is not part of a doublet.
Furthermore, processes with double Higgs boson production can be predicted to a certain extent in terms of 
single-Higgs couplings, and can thus be used to probe the nature of the Higgs boson~\cite{WP2}.

\section{Implementing the Higgs effective Lagrangian beyond the tree level}
\label{sec:beyondtree}

In this section we address a few issues related to the use of the effective Lagrangians~(\ref{eq:effL})
and (\ref{eq:chiralL}) beyond the tree level, as
required to make Higgs precision physics without assuming the validity of the Standard Model. While the methodology 
is well established and various examples of its application exist in several different contexts, 
we think that a dedicated discussion can be useful to better clarify some specific points (see also Ref.~\cite{Passarino:2012cb} for a recent discussion). 
As an illustrative though important example, we will consider the calculation of the Higgs partial decay widths, and show how the corrections from 
dimension-6 operators can  be incorporated in a consistent way.
As a by-product of our analysis and to better demonstrate its applicability,  in a companion paper~\cite{eHDECAYpaper} we will present
a modified version of the program {\tt HDECAY}~\cite{hdecay}
that features a full implementation of the effective Lagrangian $\Delta {\cal L}_{SILH}$, Eq.~(\ref{eq:silh}), as well as its generalization to the case of 
a non-linearly realized EW symmetry, Eq.~(\ref{eq:chiralL}).

A first difficulty which arises when using either Eq.~(\ref{eq:effL}) or  (\ref{eq:chiralL}) is the presence of multiple expansion parameters.
For generic values of the Higgs couplings $c_i$, the validity of the effective Lagrangian~(\ref{eq:chiralL}) is based on a double perturbative expansion in the 
SM couplings, $\alpha_{SM}/4\pi$, and in powers of $E/M$. 
The effective Lagrangian~(\ref{eq:effL}) further assumes $(v/f) \ll 1$, which implies small shifts in the Higgs couplings:
$c_i = 1 + \delta c_i$, with $\delta c_i \lesssim O(v^2/f^2)$.
All these expansion parameters must  be properly taken into account when performing calculations.
Furthermore, the non-renormalizability of the effective theory implies the presence of additional divergences compared to the SM case
which must be absorbed by a renormalization of the Wilson coefficients of local operators.

\subsection{RG evolution of the Wilson coefficients}

Let us discuss the  issue of the renormalization and RG evolution of the Wilson coefficients first.
As done in the previous sections, we will assume that the Higgs boson is part of an $SU(2)_L$ doublet and use the Lagrangian~(\ref{eq:effL}).
Since we are only interested in the divergent structure of the diagrams, it is convenient to work in the limit of unbroken $SU(2)_L\times U(1)_Y$
and compute the Green functions in terms of the Higgs doublet~$H$.
The only 1-loop diagrams which generate additional logarithmic divergences are those featuring one
insertion of the effective vertices from dimension-6 operators. By dimensional analysis, further insertions of the effective vertices lead
to power-divergent contributions to dimension-6 operators (which are irrelevant to determine the RG running)
and log-divergent contributions to higher-dimensional operators. 
The same counting holds also at higher loop level: the only log-divergent contribution to dimension-6 operators comes from
diagrams with one insertion of the effective couplings, and is thus suppressed by extra powers of the SM expansion parameter 
$\alpha_{SM}/4\pi$. This is in analogy with the renormalization of the pion effective
Lagrangian in the chiral limit, see Ref.~\cite{Weinberg:1978kz}.
It thus follows that the RG equation is linear and homogeneous in the  $\bar c_i$,
and different operators with the same quantum numbers will in general mix with each other.
At leading order in $\alpha_{SM}$, with $\alpha_{SM} = \alpha_{em}, \alpha_2, \alpha_s$, respectively, in the case of electromagnetic, weak
and QCD corrections, one has
\begin{equation}
\label{eq:RGci}
\bar c_i(\mu) = \left( \delta_{ij} +  \gamma^{(0)}_{ij}\, \frac{\alpha_{SM}(\mu)}{4\pi} \log\!\left(\frac{\mu}{M}\right) \right) \bar c_j(M) \, ,
\end{equation}
where $\gamma^{(0)}_{ij}$ is the leading-order coefficient of the anomalous dimension.
Some elements of the anomalous dimension matrix $\gamma^{(0)}_{ij}$ have been recently computed in Refs.~\cite{Grojean:2013kd, Elias-Miro:2013gya}.

In the case in which the Higgs boson and possibly the SM quarks (in particular the top and the bottom) are strongly coupled to the
new dynamics, the leading RG running effect  comes from loops of these particles and 
can be as large as $\Delta {\bar c}_i/{\bar c}_i(M) \sim (g_*^2/16\pi^2) \log(M/\mu)$ or $(\lambda_\psi^2/16\pi^2) \log(M/\mu)$.
This must be compared to the effects of order $(g_{SM}^2/16\pi^2) \log(M/\mu)$  from loops of gauge fields.
For example, the insertion of ${\bar c}_{H}$
 in the diagram $(a)$ of Fig.~\ref{fig:1loopdiagrams} leads to
a renormalization of $O_{W+B} \equiv O_W + O_B$:
\begin{equation}
\bar c_{W+B}(\mu) = \bar c_{W+B}(M)  -\frac{1}{6}\;  \frac{\alpha_2}{4\pi} \log\left(\frac{\mu}{M}\right) \bar c_H(M)\, ,  
\end{equation}
where $\alpha_2$ has been defined in Eq.~(\ref{eq:alpha2}).
It is well known that this RG running is associated with the IR contribution
to the $\epsilon_3$ parameter, and the same coefficient $\gamma^{(0)}_{W+B,H}=-1/6$ can indeed be extracted from self-energy diagrams~\cite{Barbieri:2007bh}.
%
\begin{figure}[t]
\begin{center}
\includegraphics[width=0.3\linewidth]{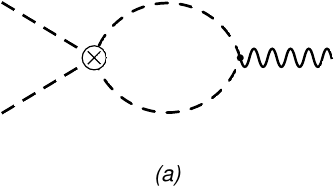}
\hspace{1.5cm}
\includegraphics[width=0.3\linewidth]{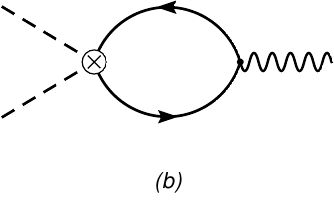}
\caption{\small 
One-loop diagrams relevant for the RG running of $\bar c_{W}$ and $\bar c_B$.
Dashed, continuous and wiggly lines denote, respectively, a weak doublet $H$, a fermion and a vector field $V=W,B$.
The symbol $\otimes$ denotes the insertion of the effective vertex from $O_H$ (in diagram~$(a)$) or $O_{H\psi}$ (in diagram~$(b)$).
}
\label{fig:1loopdiagrams}
\end{center}
\end{figure}
%
%
From the estimates of Eq.~(\ref{eq:NDAestimates}), $\bar c_H(M)\sim O(v^2/f^2)$, $\bar c_{W,B}(M) \sim O(m_W^2/M^2)$, 
it follows that the correction to $\bar c_{W+B}$ from its RG evolution down to the scale $\mu$ is of order 
$\Delta {\bar c}_{W+B}/ {\bar c}_{W+B}(M) \sim (g_*^2/16\pi^2) \log(M/\mu)$ as anticipated.
Similarly, the insertion of $\bar c_{H\psi}$ into a loop of fermions, like in  diagram~$(b)$ of Fig.~\ref{fig:1loopdiagrams}, leads to a renormalization of 
$\bar c_{W}$ and $\bar c_B$:
\begin{equation}
\Delta {\bar c}_{W,B} \approx N_c \frac{\alpha_2}{4\pi} \log\left(\frac{\mu}{M}\right)  {\bar c}_{H\psi}(M)\, ,
\end{equation}
where $N_c =3$ is a color factor.
In this case the RG correction is of order $(\lambda_\psi^2/16\pi^2) \log(M/\mu)$ compared to the UV value of the coefficients, as one can immediately verify
by using the estimates~(\ref{eq:NDAestimates}).

Loops of EW gauge fields give corrections which are suppressed by a weak loop factor $(g^2/16\pi^2)$, and the
associated RG evolution is therefore generically small.
An important exception is the case in  which the Wilson coefficient has a value  suppressed at the scale $M$.
For example, if the dynamics behind the EW symmetry breaking is custodially invariant, then $\bar c_T(M)=0$.
%
\begin{figure}[t]
\begin{center}
\includegraphics[width=0.3\linewidth]{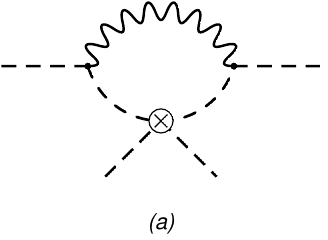}
\hspace{1.5cm}
\includegraphics[width=0.23\linewidth]{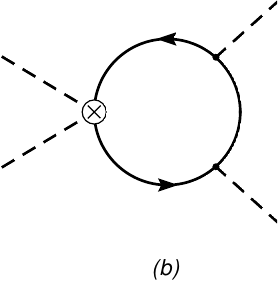}
\caption{\small 
One-loop diagrams relevant for the RG running of $\bar c_{T}$.
Dashed, continuous and wiggly lines denote, respectively, a weak doublet $H$, a fermion and a hypercharge field $B$.
The symbol $\otimes$ denotes the insertion of the effective vertex from $O_H$ (in diagram~$(a)$) or $O_{H\psi}$ (in diagram~$(b)$).
}
\label{fig:1loopdiagrams-2}
\end{center}
\end{figure}
%
%
The insertion of $\bar c_H$ into a loop of hypercharge gauge bosons, as in diagram~$(a)$ of Fig.~\ref{fig:1loopdiagrams-2}, renormalizes $\bar c_T$
and gives
\begin{equation}
\bar c_T(\mu) = \frac{3}{2} \tan^2\!\theta_W\,  \frac{\alpha_2}{4\pi} \log\left(\frac{\mu}{M}\right) \bar c_H(M)\, .
\end{equation}
Compared to the naive estimate of Eq.~(\ref{eq:NDAestimates}), $\bar
c_T(M)\sim O(v^2/f^2)$, valid in absence of custodial symmetry, 
the above correction is further suppressed by a factor $(g^{\prime\, 2}/16\pi^2) \log(M/\mu)$.
Although small, such a low-energy value of $\bar c_T$ has a strong impact on the EW precision tests performed at LEP~\cite{Barbieri:2007bh}.~\footnote{For example, 
$\bar c_T(m_Z) \sim 10^{-3}$ for $\bar c_H(M) \sim 0.1 $.}
On the other hand, it is too small to be observable through a measurement 
of the Higgs couplings at the LHC.  A similar renormalization of $\bar c_T$ also follows from loops of SM fermions through the insertion of $\bar c_{H\psi}$,
as illustrated by diagram $(b)$ of Fig.~\ref{fig:1loopdiagrams-2}. The explicit calculation for the case of a composite right- and left-handed top quark
was  performed for example in Ref.~\cite{Pomarol:2008bh}. Naively, the effect goes like
\begin{equation}
\Delta {\bar c}_{T} \approx N_c \frac{y_\psi^2}{16\pi^2} \log\left(\frac{\mu}{M}\right)  {\bar c}_{H\psi}(M)\, ,
\end{equation}
and is of order $(y_\psi/g^\prime)^2 (\lambda_\psi/g_*)^2$ compared to the one from loops of hypercharge.~\footnote{Notice that 
in case of a sizable fermion coupling $\lambda_\psi$,
a numerically larger contribution to $\bar c_T$ comes from fermionic loops with {\it two} insertions
of $\bar c_{H\psi}$. The corresponding diagram is quadratically divergent, so that it gives a  threshold correction to $\bar c_T$ at the scale $M$,
but does not contribute to its running. An explicit calculation can be found in Ref.~\cite{Pomarol:2008bh} for the case of a composite top quark.
Naively the effect is of order  $\Delta \bar c_T \sim N_c (v/f)^2 (\lambda_\psi/16\pi^2)  (\lambda_\psi/g_*)^2$, and can be numerically large.
For example, if both $t_L$ and $t_R$ couple with the same strength $\lambda_{t_L} = \lambda_{t_R}\sim \sqrt{g_* y_t}$ to the new dynamics, then it follows 
$\Delta \bar c_T \sim N_c (v/f)^2 (y_t^2/16\pi^2)$.
}

In general, although small, the RG evolution of the Wilson coefficients due to EW loops must be properly taken into account in order to precisely match the
experimental results obtained at  low energy with the theory predictions at high energy.
This is even more true in the case of QCD loop corrections, which can be large and will affect the coefficients of the dimension-6 operators
with quarks and gluon fields.~\footnote{Notice that $g_s^2 \bar c_g$ is not renormalized at one-loop by  QCD corrections. This follows from the RG-invariance of
the operator $(\beta(g_s)/g_s) G_{\mu\nu}G^{\mu\nu}$ which contributes to the trace of the energy-momentum tensor~\cite{Adler:1976zt, Collins:1976yq, Nielsen:1977sy}. See also the recent discussion
in Ref.~\cite{Grojean:2013kd}.}
The effect of the running of the Wilson coefficients can be easily incorporated in programs for the automatic calculation of production cross sections 
and decay rates by
using the effective Lagrangian~(\ref{eq:effL}) and identifying the coefficients  appearing there as their values at the relevant low-energy scale. 


\subsection{Decay rates at the loop level with the effective Lagrangian}
\label{sec:decayat1L}

In addition to the short-distance effects discussed above, 
which are parametrized in terms of the evolution of the coefficients of local operators, 
one-loop diagrams also lead to long-distance corrections to the observables under consideration.
Specifically, while short-distance effects are related to the divergent terms,  the long-distance contributions
correspond to the finite parts and are defined in a given renormalization scheme.
In general,  the decay amplitude can be expanded as follows:~\footnote{\label{footnote:EW} 
In the strict sense this equation is valid for the genuine EW corrections only, while for 
simplicity we include the (IR-divergent) virtual QED corrections to the SM amplitude in the same way. The corresponding real photon radiation contributions 
to the decay rates are treated in terms of a {\it linear} novel contribution to the Higgs coupling for the squared amplitude in order to obtain an infrared finite 
result. Pure QED corrections factorize as QCD corrections in general so that their amplitudes scale with the modified Higgs couplings. However, they cannot 
be separated from the genuine EW corrections in a simple way.}
\begin{equation}
\label{eq:ampl}
A = A^{SM}_0 + A^{SM}_1 + \Delta A_0 + \Delta A_1 + \dots
\end{equation}
where $A_0^{SM}$ ($A_1^{SM}$) is the tree-level (one-loop) SM amplitude, and
$\Delta A_0$  ($\Delta A_1$) is the  tree-level (one-loop) contribution from the  dimension-6 operators of the effective Lagrangian 
in Eqs.~(\ref{eq:silh})--(\ref{eq:silh3}). The dots denote higher-loop contributions as well as the corrections due to higher-order operators.

Let us consider for example the decay $h\to W^{(*)}W^*$.
In this case the  operators that can contribute at tree-level are $O_H$, $O_{W}$, $O_{HW}$, $O_{\psi W}$, $O_{H\psi}^\prime$, 
as well as $O_{Hud}$ in the case in which the off-shell $W$ decays into a pair of quarks. 
Based on the naive estimates of Eq.~(\ref{eq:NDAestimates}) and according 
to the discussion of Section~\ref{sec:NPvsSM}, we can quantify the various effects encoded by $\Delta A_0$ as follows:
\begin{equation}
\label{eq:WWanatomy}
\begin{split}
\frac{\Delta A_0}{A_0^{SM}}(W^{(*)}W^*) = 
& \, \hat c_H\times O\!\left(\frac{v^2}{f^2}\right) +\hat c_W\times O\!\left(\frac{E^2}{M^2}\right) + \hat c_{HW}\times O\!\left(\frac{E^2}{16\pi^2 f^2}\right)  
\\[0.2cm]
& + \hat c_{Hud}\times O\!\left(\frac{v^2}{f^2} \, \frac{\lambda_u \lambda_d}{g_*^2} \right)  
   + \hat c_{H\psi}^\prime\times O\!\left(\frac{v^2}{f^2} \, \frac{\lambda_\psi^2}{g_*^2} \right)  
   + \hat c_{\psi W}\times O\!\left(\frac{E m_\psi}{16\pi^2 f^2} \right)  \, .
\end{split}
\end{equation}
Here $E= m_h$ is the relevant energy of the process and we have conveniently defined each of the $O(1)$ parameters $\hat c_i$ 
to be equal to $\bar c_i(m_h)$  divided by its naive estimate in Eq.~(\ref{eq:NDAestimates}):
\begin{equation}
\label{eq:hatc}
\begin{gathered}
 \hat c_i = \frac{f^2}{v^2} \bar c_i(m_h), \quad  i=H, T, 6, \psi, \qquad\quad
\hat c_i = \frac{M^2}{m_W^2} \bar c_i(m_h), \quad i=W,B\, ,\\[0.2cm]
\hat c_i = \frac{16 \pi^2 f^2}{m_W^2} \bar c_i(m_h), \quad i = HW, HB, \gamma, g, \psi W, \psi B, \psi G \, ,\\[0.2cm]
\hat c_i = \frac{g_*^2}{\lambda_\psi^2}\frac{f^2}{v^2} \bar c_i(m_h), \quad
\hat c_i^\prime = \frac{g_*^2}{\lambda_\psi^2}\frac{f^2}{v^2} \bar c_i^\prime(m_h), \quad i=H\psi \, ,
\qquad\quad
\hat c_{Hud} = \frac{g_*^2}{\lambda_u\lambda_d}\frac{f^2}{v^2} \bar c_{Hud}(m_h)\, .
\end{gathered}
\end{equation}
When the Higgs boson is pNGB, the two parameters $\hat c_{g}$ and $\hat c_\gamma$ are not of order one but are further suppressed by a factor $g_{\not G}^2/g_*^2$.
From Eq.~(\ref{eq:WWanatomy}) one can see that the contribution of the dipole operators $O_{\psi W}$ is suppressed by $(m_\psi/m_h)$
compared to that of $O_{HW}$, while that of 
$O_{Hud}$ and $O_{H\psi}^\prime$ is expected to be small given the existing constraints on the couplings 
$\lambda_\psi$ (see the discussion in Section~\ref{sec:bounds}).
The dominant NP contribution thus comes from the terms in the first line of Eq.~(\ref{eq:WWanatomy}), among which the one proportional to
$\bar c_H$ is  the leading effect for $g_* > g$. The 1-loop electroweak amplitude $A_1^{SM}$ gives a contribution of order 
$A_1^{SM}/A_0^{SM} \sim (\alpha_2/4\pi)$.
We thus see explicitly that $\Delta A_0$ and $A_1^{SM}$ encode the NLO corrections in the three expansion parameters which we are considering: 
$\alpha_2/4\pi$ (electroweak expansion), $E^2/M^2$ (derivative expansion) and $v^2/f^2$.
The  contribution  due to 1-loop diagrams with one insertion of the effective vertices  has
not been computed yet, but we can easily estimate its size:
\begin{equation}
\label{eq:WWanatomy1L}
\frac{\Delta A_1}{A_0^{SM}}(W^{(*)}W^*) = 
\hat c_H\times O\!\left(\frac{v^2}{f^2} \, \frac{\alpha_2}{4\pi}\right) + \hat c_u\times O\!\left(\frac{v^2}{f^2} \, \frac{\alpha_2}{4\pi}\right) 
+ \hat c_6\times O\!\left(\frac{v^2}{f^2} \, \frac{\alpha_2}{4\pi}\right) 
+ \dots
\end{equation}
where the dots denote the subleading terms due to the other operators.
The terms shown in Eq.~(\ref{eq:WWanatomy1L}) arise from the same 1-loop diagrams 
that give the SM amplitude $A_1^{SM}$, where each of the Higgs couplings gets shifted by $\bar c_H$, $\bar c_u$ and $\bar c_6$.
By neglecting the unknown $\Delta A_1$ one is omitting terms of order $(v^2/f^2) (\alpha_2/4\pi)$, that is, of the same size of the 
tree-level contribution due to the operator $O_{HW}$, see Eq.~(\ref{eq:WWanatomy}), since $E = m_h \approx m_W$. This latter contribution can
be easily computed and it is included in the formula of the decay rate to $WW$ (and similarly that of $O_{HW}$ and $O_{HB}$ to $ZZ$ is also included) 
implemented in the program \texttt{eHDECAY} discussed in Ref.~\cite{eHDECAYpaper}.
%
%
The addition of the tree-level correction from $O_{HW}$ is clearly
the first step towards a full inclusion of the $O[(v^2/f^2) (\alpha_2/4\pi)]$ corrections, where the missing part will have to be computed
from 1-loop diagrams featuring one insertion of $O_H$, $O_u$ and $O_6$. It is worth noting that these diagrams, in general, contain logarithmic
divergences which must be reabsorbed by a renormalization of the Wilson coefficients and contribute to their RG evolution as explained in the previous
section. The finite part is the contribution to $\Delta A_1$ which awaits to be computed.

By approximating the amplitude as $A \simeq A_0^{SM} + A_1^{SM} + \Delta A_0$ one obtains
the following  formula for the decay rate:~\footnote{The same remark as in footnote~\ref{footnote:EW} applies.}
\begin{equation}
\label{eq:WWrateNLO}
\begin{split}
\Gamma(W^{(*)}W^*) = \Gamma_0^{SM}(W^{(*)}W^*)  \Bigg\{  
& 1 + \frac{2}{|A_0^{SM}|^2}\, \text{Re}\!\left[ \left(A_0^{SM}\right)^*\left( A_1^{SM} +\Delta A_0  \right) \right] \\[0.1cm]
& + O\!\left(  \left( \frac{v^2}{f^2}\right)^2 , \left(\frac{\alpha_2}{4\pi} \, \frac{v^2}{f^2}\right) , \left(\frac{\alpha_2}{4\pi}\right)^2 \right) 
\Bigg\}\, ,
\end{split}
\end{equation}
where $\Gamma_0^{SM}(W^{(*)}W^*)$ denotes the tree-level SM decay rate.
For simplicity,  we have not shown  terms  involving powers of $E^2/M^2$ among the neglected
contributions, since for $E=m_h \approx m_W$ one has $E^2/M^2 \lesssim v^2/f^2$ if $g_* \gtrsim g$.
As mentioned, this formula incorporates the $O(v^2/f^2)$, $O(\alpha_2/4\pi)$ and $O(m_h^2/M^2)$ corrections (NLO in the perturbative expansion),
and can be easily implemented in existing codes for the automatic computation of the decay rate. 
The inclusion of the $O(m_h^2/M^2)$ tree-level correction due to $O_W$ is justified as long as $g_* < 4\pi$, since it is parametrically larger
 than the neglected $O[(v^2/f^2) (\alpha_2/4\pi)]$ terms by a factor $(16\pi^2/g_*^2)$.
Notice that in the limit of large deviations of the Higgs couplings from their SM values, $(v/f)^2 \sim O(1)$, the neglected terms of $O[(v^2/f^2) (\alpha_2/4\pi)]$
become as important as those included through $A_1^{SM}$. In order words, a proper inclusion of the EW corrections in the limit $v\sim f$
requires a complete 1-loop calculation where each of the diagrams is rescaled by the appropriate coupling factor. 

A similar discussion  applies to the Higgs decay into a pair of fermions, $h\to \bar \psi \psi$. 
In this case only $O_H$ and $O_\psi$ ($\psi = u,d,l$)
contribute at tree level,
\begin{equation}
\frac{\Delta A_0}{A_0^{SM}}(\bar\psi\psi) =  \left(\frac{\hat c_H}{2} +  \hat c_\psi \right)\times O\!\left(\frac{v^2}{f^2}\right)  \, ,
\end{equation}
while the one-loop EW diagrams featuring one effective vertex give a correction 
of order
\begin{equation}
\frac{\Delta A_1}{A_0^{SM}}(\bar\psi\psi) = 
\hat c_H\times O\!\left(\frac{v^2}{f^2} \, \frac{\alpha_2}{4\pi}\right) + \hat c_\psi\times O\!\left(\frac{v^2}{f^2} \, \frac{\alpha_2}{4\pi}\right) 
+ \hat c_6\times O\!\left(\frac{v^2}{f^2} \, \frac{\alpha_2}{4\pi}\right) + \dots
\end{equation}
where the dots indicate the subleading terms due to the other operators.
The calculation of $\Delta A_1$ has not been performed yet, while the 1-loop EW corrections are known in the SM, $A_1^{SM}$.
Their inclusion is thus possible as long as $(v/f) \ll 1$, so that the neglected terms in $\Delta A_1$ are subleading.
The case of QCD radiative corrections is different, since at leading order they factorize with respect to the expansion in the number of derivative
and  fields and can thus be resummed up to higher orders.  In the case of the Higgs decay into a pair of quarks one can for example approximate 
$A \simeq A_0^{SM} + A_1^{SM} + \Delta A_0$ and obtain the following formula for the decay 
rate:~\footnote{The same remark as in footnote~\ref{footnote:EW} applies.}
\begin{equation}
\label{eq:qqrateNLO}
\begin{split}
\Gamma(\bar q q) = \Gamma_0^{SM}(\bar q q)  \, \kappa^{QCD} \Bigg\{  
& 1 + \frac{2}{|A_0^{SM}|^2}\, \text{Re}\!\left[ \left(A_0^{SM}\right)^*\left( A_1^{SM} +\Delta A_0  \right) \right] \\[0.1cm]
& + O\!\left(  \left( \frac{v^2}{f^2}\right)^2 , \left(\frac{\alpha_2}{4\pi} \, \frac{v^2}{f^2}\right) , \left(\frac{\alpha_2}{4\pi}\right)^2 \right) 
\Bigg\}\, ,
\end{split}
\end{equation}
where $\Gamma_0^{SM}(\bar q q)$ is the SM tree-level rate and $\kappa^{QCD}$ encodes the QCD corrections. 
This formula includes the leading $O(v^2/f^2)$, $O(\alpha_2/4\pi)$ and QCD corrections. Mixed electroweak and QCD corrections 
can also be included by assuming that they factorize, as the non-factorizable terms are known to be small. 
Compared to the decay rate into $WW$,  Eq.~(\ref{eq:qqrateNLO}) apparently does not include corrections of order $m_h^2/M^2$. While there
is indeed no operator whose contribution starts at that order,  such corrections can arise from subleading contributions to 
$\bar c_H$ and $\bar c_\psi$. For example, the tree-level exchange of heavy fermions can lead to a wave-function renormalization of the SM 
ones, which  can be re-expressed  in our notation as a contribution to $\bar c_\psi$ of order $\lambda_\psi^2 v^2/M^2$.

A similar resummation of the QCD corrections also works for the decay  $h\to gg$.
In this case the SM tree-level amplitude vanishes,  $A^{SM}_0=0$, while the leading contribution arises from the 1-loop exchange of top quarks.
The two-loop EW corrections are known in the SM and give a correction of order $A_2^{SM}/A_1^{SM} \sim \alpha_2/4\pi$.
Among the dimension-6 operators, only $O_{g}$ contributes at tree-level,
\begin{equation}
\frac{\Delta A_0}{A_1^{SM}}(gg) = \hat c_g \times O\!\left(\frac{v^2}{f^2}\right)\, . 
\end{equation}
As discussed in Section~\ref{sec:effLag} 
(see Eq.~(\ref{eq:cgcga})),  the above estimate is suppressed by an additional factor $(g^2_{\not G}/g_*^2)$ in the case of a NG Higgs boson.
At the one-loop level one has
\begin{equation}
\frac{\Delta A_1}{A_1^{SM}}(gg) = \left(\frac{\hat c_H}{2} + \hat c_u \right) \times O\!\left(\frac{v^2}{f^2}\right)
 + \hat c_{tG} \times  O\!\left(\frac{v^2}{f^2}\, \frac{y_t^2}{16\pi^2} \right) \, .
\end{equation}
Thus, the one-loop effect of $O_H$ and $O_u$ is expected to be as important as the 
tree-level one from $O_g$, and even larger if the Higgs is a NG boson,
as discussed in Section~\ref{sec:op1L}. This is in fact not surprising, since $\bar c_g$ arises at the 1-loop level
in minimally coupled theories, while $\bar c_H$ and $\bar c_u$ can be generated at tree level.
The contribution from the dipole operator $O_{tG}$ is suppressed by a factor $y_t^2/16\pi^2$ compared to that
from $O_H$ and $O_u$, as expected from the fact that $\bar c_{tG}$
is generated at the 1-loop level in minimally coupled theories. For this reason  it can be neglected.
%
It should be noted that without a complete computation of the NLO EW corrections of order $(\alpha_2/4\pi) (v^2/f^2)$, the LHC data on Higgs physics 
are not sensitive to the range of values of $\bar c_{tG}$ expected using the  naive estimate (\ref{eq:NDAestimates}) with $(v/f)^2\sim 0.1$.
Furthermore, we stress that in order to distinguish the effect of $O_{tG}$ from that of $O_g$, the $t\bar t h$ channel should be measured~\cite{Degrande:2012gr} 
(single top production in association with the Higgs could also provide complementary information~\cite{Farina:2012xp}).
Also in this case, there are no operators giving $m_h^2/M^2$ corrections, although these terms will in general appear as subleading
contributions to $\bar c_{g}$, $\bar c_H$ and $\bar c_u$, as discussed above.
It is well known that  higher-order $\alpha_s$ corrections are large, so 
they must be  included consistently in our perturbative expansion. This can be  done easily in the approximation $m_h \ll 2 m_t$, which is
reasonably accurate for $m_h = 125\,$GeV. In such a limit one can integrate out the top quark and match its one-loop contribution to 
that of the local operator $O_g$. Then it trivially follows that the QCD  corrections associated to the virtual exchange and real emissions 
of gluons and light quarks below the scale $m_t$ factorize in the rate, the multiplicative factor being the same for both the top quark and
New Physics terms. By approximating $A \simeq A_1^{SM} + A_2^{SM} + \Delta A_0 + \Delta A_1$,
one arrives at the following  formula for the $h\to gg$ decay rate:
\begin{equation}
\label{eq:ggrate}
\begin{split}
\Gamma(gg) = \Gamma_1^{SM}(gg) \, \kappa_{soft} \, \Bigg\{ 
&  c_{eff}^2 + \frac{2\, c_{eff} }{|A_1^{SM}|^2}\, \text{Re}\!\left[ \left(A_1^{SM}\right)^* 
\left( A_2^{SM} c_{eff} + \Delta A_0 + \Delta A_1 \, c_{eff} \right) \right] \\[0.1cm]
& + O\!\left(  \left( \frac{v^2}{f^2}\right)^2 , \left(\frac{\alpha_2}{4\pi} \, \frac{v^2}{f^2}\right) , \left(\frac{\alpha_2}{4\pi}\right)^2 \right) 
\Bigg\}, 
\end{split}
\end{equation}
where $\Gamma_1^{SM}(gg)$ is the 1-loop SM decay width.
The factor $c_{eff}$ includes all the dependence on $m_t$ and accounts for virtual QCD corrections to $A^{SM}_1$ above that scale,
while $\kappa_{soft}$ parametrizes the soft radiative effects. By using Eq.~(\ref{eq:ggrate}), the existing four-loop calculations of 
$c_{eff}$~\cite{Chetyrkin:1997un,Kramer:1996iq,Schroder:2005hy,Chetyrkin:2005ia}
and $\kappa_{soft}$~\cite{nloggqcd,Spira:1995rr,Baikov:2006ch} 
allow one to include the QCD corrections up to~N$^3$LO. 

The contributions  to the decay $h\to \gamma\gamma$ follow a similar pattern as for $h\to gg$.
At tree level:
\begin{equation}
\frac{\Delta A_0}{A_1^{SM}}(\gamma\gamma) = \hat c_\gamma \times O\!\left(\frac{v^2}{f^2}\right)\, .
\end{equation}
At one loop:
\begin{equation}
\label{eq:gagaanatomy1L}
\begin{split}
\frac{\Delta A_1}{A_1^{SM}}(\gamma\gamma) = 
& \, \hat c_H \times O\!\left(\frac{v^2}{f^2}\right) + \hat c_u \times O\!\left(\frac{v^2}{f^2}\right) +\hat c_W\times O\!\left(\frac{m_W^2}{M^2}\right) 
\\[0.2cm]
& + \hat c_{HW}\times O\!\left(\frac{m_W^2}{16\pi^2 f^2}\right) 
    + ( \hat c_{tW} + \hat c_{tB}) \times  O\!\left(\frac{v^2}{f^2}\, \frac{y_t^2}{16\pi^2} \right)  \, . 
\end{split}
\end{equation}
The 2-loop electroweak corrections have been computed in the SM and can be included for $(v^2/f^2)\ll 1$, so that unknown $O[(v^2/f^2)(\alpha_2/4\pi)]$
effects arising from 2-loop diagrams with one effective vertex are negligible. 
From Eq.~(\ref{eq:gagaanatomy1L}) one can see that the 1-loop contribution due to $O_{HW}$ is of the same order as such neglected terms.
%
%
The 1-loop correction from $O_W$, on the contrary,  is parametrically larger  by a factor $(16\pi^2/g_*^2)$
and should be included for $g_* < 4\pi$. The easiest way to compute it is by rewriting $O_W$ in terms of the other operators through the equations
of motions~\cite{SILH}, see Eq.~(\ref{eq:OWrewritten2}). 
%
%
The 1-loop correction due to the dipole operators is suppressed by a factor $y_t^2/16\pi^2$ and can be neglected.
Approximating $A \simeq A_1^{SM} + A_2^{SM} + \Delta A_0 + \Delta A_1$ one finds:
\begin{equation}
\label{eq:gagarate}
\begin{split}
\Gamma(\gamma\gamma) = \Gamma_1^{SM}(\gamma\gamma)\,  \Bigg\{ 
&  1 + \frac{2}{|A_1^{SM}|^2}\, \text{Re}\!\left[ \left(A_1^{SM}\right)^* 
\left( A_2^{SM} + \Delta A_0 + \Delta A_1 \right) \right] \\[0.1cm]
& + O\!\left(  \left( \frac{v^2}{f^2}\right)^2 , \left(\frac{\alpha_2}{4\pi} \, \frac{v^2}{f^2}\right) , \left(\frac{\alpha_2}{4\pi}\right)^2 \right) 
\Bigg\}, 
\end{split}
\end{equation}

Finally, the estimate of the corrections to $h\to \gamma Z$ is the following:
\begin{align}
\frac{\Delta A_0}{A_1^{SM}}(Z\gamma) = 
& \,\hat c_\gamma \!\times\! O\!\left(\frac{v^2}{f^2}\right) + (\hat c_{HW} - \hat c_{HB} ) \!\times\! O\!\left(\frac{v^2}{f^2}\right) \, , 
\\[0.5cm]
\begin{split}
\frac{\Delta A_1}{A_1^{SM}}(Z\gamma) = 
& \, \hat c_H \!\times\! O\!\left(\frac{v^2}{f^2}\right) + \hat c_u \!\times\! O\!\left(\frac{v^2}{f^2}\right) +\hat c_W \!\times\! O\!\left(\frac{m_W^2}{M^2}\right) 
   + \hat c_{HW} \! \times \! O\!\left(\frac{m_W^2}{16\pi^2 f^2}\right) 
    \\[0.2cm]
& + \hat c_{tW} \!\times\!  O\!\left(\frac{v^2}{f^2}\, \frac{y_t^2}{16\pi^2} \right) 
   + \hat c_{tB}  \!\times\!  O\!\left(\frac{v^2}{f^2}\, \frac{y_t^2}{16\pi^2} \right)  \, . 
\end{split}
\end{align}
In this case the 1-loop electroweak corrections are not known in the SM, so that the formula for the decay rate reads:
\begin{equation}
\Gamma(Z\gamma) = \Gamma_1^{SM}(Z\gamma)\,  \Bigg\{ 
 1 + \frac{2}{|A_1^{SM}|^2}\, \text{Re}\!\left[ \left(A_1^{SM}\right)^* 
\left( \Delta A_0 + \Delta A_1 \right) \right] 
+ O\!\left(  \left( \frac{v^2}{f^2}\right)^2 , \left(\frac{\alpha_2}{4\pi} \right) \right) 
\!\Bigg\}, 
\end{equation}
where only the contributions from $O_H$, $O_u$ and $O_W$ should be retained in $\Delta A_1$ for consistency.

\vspace{0.5cm}
Through the above discussion we sketched  how the effective Lagrangian can be implemented beyond the tree level in the calculation
of physical quantities. In the case of the Higgs partial decay widths, in particular, we have seen how the EW and QCD corrections can be
included consistently with the expansion in the number of fields and derivatives.
As a more concrete illustration of these considerations, we have written a modified version of the program {\tt HDECAY}, which we
dub {\tt eHDECAY}, where the corrections from all the local operators of the effective Lagrangians~(\ref{eq:silh}) and~(\ref{eq:chiralL}) are included at  NLO.
A detailed description of the code is given in Ref.~\cite{eHDECAYpaper},
where more explicit formulas for each of the Higgs partial widths are provided.

\section{Discussion}
\label{sec:conclusions}

The discovery of a resonance with a mass around 125~GeV similar to the long-sought Standard Model Higgs boson brings the exploration of the electroweak 
symmetry breaking sector under quantitative scrutiny. The LHC experiments, together with those at the Tevatron, report the signal strengths, i.e. the product of 
the Higgs production cross section times its decay branching ratio, for various final state channels. The main task of the community is now to interpret these 
data and understand the implications for the theory of New Physics that is expected to lie beyond the weak scale.

The EW oblique parameters provide a bound on the scale of New Physics but do not give detailed information about the nature of the NP sector. 
In order to understand how the weak scale is stabilized at the quantum level, i.e. how the hierarchy problem is solved, one crucial question is whether 
EW symmetry breaking proceeds by weak or  strong dynamics. 
The direct observation of new degrees of freedom would provide a  straightforward answer. But a glimpse of New Physics can also be caught from a dedicated study of the Higgs boson itself, and in particular from a measurement of its couplings, if a departure from the SM predictions is ever observed. It is useful to parametrize the deviations from the SM by the effective Lagrangian of Eq.~(\ref{eq:effL}).  By measuring its Wilson coefficients $\bar c_i$ one can infer what kind of UV theory
completes the SM.

If the coupling strength of the Higgs boson to the NP sector is of the  order of the SM weak couplings, $g_* \approx g$,
then our power counting~(\ref{eq:NDAestimates}) shows that the  coefficients of  the operators  that can be generated at tree-level,
$O_H, O_{u,d,l}, O_W$ and $O_B$, 
are  expected to be all of the same order, $m_W^2/M^2$, where $M$ is the typical mass scale of the NP spectrum, unless some special selection rule suppresses 
some of them. It is instructive to examine the predictions of the archetypal example of weakly-coupled UV completions: the Minimal Supersymmetric
Standard Model (MSSM).
First,  $R$-parity protects the EW oblique parameters from any tree-level contributions, hence $\bar c_W$ and $\bar c_B$ are of order 
$(m^2_W/M^2) (\alpha_2/4\pi)$ 
and thus small.
Second, the couplings of the lightest Higgs boson to the massive gauge bosons
are given by $c_V=\sin(\beta-\alpha)$, where $\alpha$ is the rotation
angle to diagonalize the CP-even mass matrix and $\tan \beta$ is the ratio of the vacuum expectation values of the two neutral CP-even Higgs bosons. In the decoupling limit, $\alpha \to \beta -\pi/2$, one has
$c_V  = 1 +  O(m_Z^4/m_H^4)$, where $m_H$ is the mass of the heaviest CP-even scalar (for a general treatment 
of the decoupling limit see for example Ref.~\cite{Gunion:2002zf}).
This means that at tree-level the deviations of the Higgs-gauge boson couplings are generated by dimension-8 operators~\cite{Randall:2007as}, 
while $\bar c_H$  arises only through loop effects and is naively of order $(m^2_W/M^2) (\alpha_2/4\pi)$.
At the same time, the couplings to up- and down-type quarks read, respectively,
\begin{equation}
\begin{split}
 c_u = & + \frac{\cos \alpha}{\sin\beta} = 1+ 2\, \frac{m_Z^2}{m_H^2}\cos^2\!\beta \cos 2 \beta + O\!\left(  \frac{m_Z^4}{m_H^4} \right)\\[0.2cm]
 c_d=  & -\frac{\sin\alpha}{\cos \beta} = 1- 2\, \frac{m_Z^2}{m_H^2}\sin^2\!\beta \cos 2 \beta + O\!\left(  \frac{m_Z^4}{m_H^4} \right)\, .
 \end{split}
 \end{equation}
For moderately large $\tan\beta$ this implies $\bar c_d \sim m_Z^2/m_H^2$, while $\bar c_u$ is further suppressed by a 
factor $\sim 1/\tan^2\beta$ (see for example Refs.~\cite{Gunion:1984yn,Blum:2012ii}
and the recent discussion in Ref.~\cite{Azatov:2012qz}).
A pattern with small values of $\bar c_H, \bar c_W$, $\bar c_B$ and $\bar c_u$ but with 
a $\sim 15\%$ enhancement of the Higgs coupling to down-type quarks due to $\bar c_d$, for example,
would be indicative of the MSSM with large $\tan\beta$ and the additional Higgs bosons around $300\,$GeV.
Generic two-Higgs doublet models lead to a similar pattern of couplings, while models where the Higgs boson mixes with a  scalar that is singlet under the SM gauge group
can generate $\bar c_H$ at the tree level.
In the MSSM, loops of light stops or staus as well as charginos can also give sizable contributions to the 
effective couplings of the light Higgs boson to photons and gluons,  with $\bar c_g$, $\bar c_\gamma$ satisfying the naive estimates~(\ref{eq:NDAestimates}).
For example, loops of stops lead to $\bar c_g \sim (g_*^2/16\pi^2) (m_W^2/m_{\tilde t}^2)$, where $g_* = y_t$ or $A_t/m_{\tilde t}$.

This situation has to be contrasted with the case of strongly coupled theories. There, our power counting~(\ref{eq:NDAestimates}) singles out $\bar c_H$, 
$\bar c_{u,d}$ as the dominant Wilson coefficients ($\bar c_6$ controls only the Higgs self-interaction and measuring it at the LHC will be challenging),
while $\bar c_W$ and $\bar c_B$ are suppressed by the ratio $(g/g_{*})^2$.
Furthermore, a composite Higgs boson can be naturally light if it is the pseudo Nambu-Goldstone boson associated to the dynamical breaking of a global 
symmetry 
of the strong dynamics. This implies that the  coefficients $\bar c_g$ and $\bar c_\gamma$ will also be suppressed by a factor $(g_{\not G}/g_*)^2$, where $g_{\not G}$
is some weak spurion breaking the Goldstone symmetry. The modifications in the gluon-fusion production cross section and in the decay rate to photons 
are thus controlled by $\bar c_H$ and $\bar c_u$.

The harvest of data collected by the LHC certainly  calls for a definite theoretical framework to describe the Higgs-like resonance and  compute production and 
decay rates accurately in perturbation theory without restricting to the SM hypothesis. Effective Lagrangians are one of the tools at our disposal to achieve this 
goal. Elaborating on the operator classification of Ref.~\cite{SILH}, we  estimated the present bounds on the Wilson coefficients
and  provided accurate expressions for the Higgs decay rates including various effects that were previously omitted in the literature. 
Assuming that the observed Higgs-like resonance is a spin-0 and CP-even particle, we discussed two general formulations of the  
effective Lagrangian, one of which relies on the linear realization of $SU(2)_L\times U(1)_Y$ at high energies.
One of the questions that can be addressed by considering these two parametrizations is whether the theory of New Physics flows to the SM in the infrared, 
that is, whether the Higgs-like resonance is part of an EW doublet. 
If all the Higgs signal strengths measured at the LHC converge towards the SM prediction, it would  be a very suggestive indication that indeed the 
Higgs boson combines together with the longitudinal components of the $W$ and $Z$ to form an EW doublet, since any other alternative requires some 
tuning to fake the SM rates. On the other hand, the doublet nature of the Higgs boson would be less obvious to establish if the signal strengths exhibit deviations 
from their SM predictions (but note that some deviations in the signal strengths could unambiguously indicate that the Higgs boson is not part of a doublet, this is in particular the case if a large breaking of the custodial symmetry is observed in conflict with the strong bound already existing from EW precision data).
We have pointed out that, if the EWSB dynamics is custodially symmetric, it is not possible to test whether the Higgs boson is part of
a doublet by means of single-Higgs processes alone. A direct proof can come only from processes with multi-Higgs bosons in the final states~\cite{WP2},
which are however challenging to study at the LHC.
Precisely establishing the CP nature of the Higgs boson is another question that  also requires accurate computations. 
If there is little doubt that the observed resonance has a large CP-even component, the possibility of a small mixing with a CP-odd component remains alive, 
and dedicated analyses will have to be performed to bound the mixing angle between the two components. 
To this aim too, an effective Lagrangian including the CP-odd operators listed in Appendix~\ref{app:CPodd} provides the  theoretical framework where 
this question can be addressed quantitatively.

The absence so far  of direct signals of New Physics at the LHC  
indicates that the road to unveil the origin of the electroweak symmetry breaking
might be long and go through precision analyses rather than copious production of new particles.
For such a task, the well established technology of effective field theories is the most powerful and general tool we have to analyze the Higgs data and 
put them into a coherent picture together with the existing experimental information without assuming the validity of the Standard Model.
There is still time for the LHC to disprove this pessimistic eventuality by reporting the discovery of new light particles or large shifts in  some
of the Higgs couplings. It is clear, however,  that if the New Physics continues to remain elusive, a precise investigation of the Higgs properties will become 
the most urgent programme in high-energy physics both for  the experimental and the theoretical community.

\section*{Acknowledgments}

We thank B.~Gavela, A.~De Rujula, J.R.~Espinosa, A.~Falkowski, E.~Franco, L.~Merlo, A.~Pomarol, R. Rattazzi, F.~Riva, L.~Silvestrini, M.~Trott for insightful discussions,
and the participants of the LHC Higgs XS Working Group, in particular A.~David, A.~Denner, M.~D\"uhrssen, M.~Grazzini,  G.~Passarino
and G.~Weiglein for discussions and comments.
We also thank J.F.~Kamenik for useful explanations on the results of Ref.~\cite{Kamenik:2011dk} and we thank A.~Pomarol and E.~Masso for pointing out a sign error in  Eq.~(\ref{eq:deltag}) and we thank F.~Maltoni for reminding us about the Bianchi identities.
This research has been partly supported by the European Commission under the ERC Advanced Grant 226371 MassTeV and the contract 
PITN-GA-2009-237920 UNILHC. C.G. is  supported by the Spanish Ministry MICNN under contract FPA2010-17747. 
The work of R.C. was partly supported by the ERC Advanced Grant No. 267985 
\textit{Electroweak Symmetry Breaking, Flavour and Dark Matter: One Solution for Three Mysteries (DaMeSyFla)}.
M.M. is supported by the DFG SFB/TR9 Computational Particle Physics.

\appendix

\section{SM Lagrangian: notations and conventions}
\label{app:SM}

In this Appendix, we collect the conventions used throughout this paper.
The field content decomposes under $SU(3)_C\times SU(2)_L \times U(1)_Y$ as
\begin{eqnarray}
H=(1,2,1/2),\ L_L^i=(1,2,-1/2), \ l_R^i=(1,1,-1), \\
q_L^i=(3,2,1/6), \ u_R^i=(3,1,2/3), \ d_R^i=(3,1,-1/3),
\end{eqnarray}
where the hypercharge is defined as $Y=Q-T_{3L}$, and $i=1,2,3$ is a flavor index. 
The action of the gauge group is fully characterized by the conventions used to define the covariant derivative. For instance, for the left-handed quark doublet, we have
\begin{eqnarray}
D_\mu q_L = \left( \partial_\mu - \frac{i}{2} g_S \lambda^a g^a_\mu -  \frac{i}{2} g \sigma^i W^i_\mu - \frac{i}{6} g' B_\mu\right) q_L
\end{eqnarray}
where $\lambda^a, a=1\ldots 8$, and $\sigma^i, i=1\ldots 3$, are the usual Gell-Mann and Pauli matrices.
Accordingly, the gauge-field strengths are defined as
\begin{equation}
G_{\mu\nu}^a = \partial_\mu g_\nu^a -  \partial_\nu g_\mu^a + g_S f^{abc} g_\mu^b g_\nu^c \, ,
\end{equation}
where $f^{abc}$ are the $SU(3)$ structure constants.

The Yukawa interactions of the up-type quarks involve the Higgs charge-conjugate doublet defined as
\begin{equation}
H^c = i \sigma^2 H^*.
\end{equation}

The renormalizable Lagrangian of the SM thus reads:
\begin{equation}
\begin{split}
\mathcal{L}_{SM} = \, & -\frac{1}{4} G_{\mu\nu}^a G^{a\, \mu\nu} - \frac{1}{4} W_{\mu\nu}^i W^{i \mu\nu} -\frac{1}{4} B_{\mu\nu}B^{\mu\nu} 
+ \left(D^\mu H\right)^\dagger \!\left(D_\mu H\right) \\
&+ i\left ( {\bar L}_L\gamma^\mu D_\mu L_L + {\bar l}_R \gamma^\mu D_\mu
l_R + {\bar q}_L \gamma^\mu D_\mu q_L + {\bar u}_R \gamma^\mu D_\mu u_R + {\bar d}_R \gamma^\mu D_\mu d_R\right)\\
& + \mu_H^2 H^\dagger H - \lambda (H^\dagger H)^2 
+ (y_u \, {\bar q}_L H^c u_R + y_d \, {\bar q}_L H d_R + y_l  \, {\bar L}_L H l_R +   {\it h.c.})
\end{split}
\end{equation}

\section{Electroweak Chiral Lagrangian in non-unitary gauge}
\label{sec:EWchiralL}

We report here the expression of the EW chiral Lagrangian valid in a generic gauge and in the most general case 
in which the $SU(2)_L \times U(1)_Y$ is non-linearly realized. For simplicity, we will restrict to the case in which
the EWSB dynamics has a custodial invariance.
The scalar $h$ is assumed to be CP-even and a singlet of the custodial symmetry, and  does not necessarily
belong to an $SU(2)_L$ doublet. The Lagrangian can be expanded in terms with an increasing number of derivatives 
\begin{equation}
\label{eq:chiralLforHiggs}
{\cal L} = {\cal L}_{0} + {\cal L}_{EWSB} \, , \qquad\quad
{\cal L}_{EWSB} =   - V(h)  + {\cal L}^{(2)} +  {\cal L}^{(4)} + \dots
\end{equation}
where ${\cal L}_0$ contains the kinetic terms of the $SU(3)_c\times SU(2)_L\times U(1)_Y$  gauge fields and of the SM fermions,
${\cal L}_{EWSB}$ describes the sector responsible for  EWSB, and  $V(h)$ is the potential for  $h$~\cite{Contino:2010mh}:
\begin{equation}
\label{eq:V}
V(h) = \frac{1}{2} m_h^2 h^2 + c_3 \, \frac{1}{6} \left(\frac{3 m_h^2}{v}\right) h^3 + \dots
\end{equation}
Under the request of $SU(2)_V$ custodial symmetry,
the longitudinal $W$ and $Z$ polarizations correspond to the NG bosons of the global coset 
$SU(2)_L\times SU(2)_R/SU(2)_V$ and are described by the $2\times 2$ matrix
\begin{equation}
\Sigma(x) = \exp\left( i \sigma^a \chi^a(x)/v \right)\, ,
\end{equation}
where $\sigma^a$ are the Pauli matrices. $SU(2)_L \times U(1)_Y$ (local)
transformations read as
\begin{equation}
\Sigma(x) \to  U_L \Sigma(x) U_Y^\dagger\, , 
\qquad
U_L = \exp(i \alpha^a_L \sigma^a)\, ,  \quad
U_Y = \exp(i \alpha_Y \sigma^3)
\end{equation}
and the covariant derivative is defined by
\begin{equation}
D_\mu \Sigma = \partial_\mu \Sigma - \frac{i\, g}{2} W_\mu^a\,  \sigma^a \, \Sigma + \frac{i\, g'}{2} B_\mu\,  \Sigma\,  \sigma^3\, .
\end{equation}
At the level of two derivatives one has~\cite{Contino:2010mh}:
\begin{equation}
\label{eq:L2}
\begin{split}
{\cal L}^{(2)} =
&  \, \frac{1}{2} (\partial_\mu h)^2  + \frac{v^2}{4} \Tr\left( D_\mu \Sigma^\dagger D^\mu \Sigma \right)
    \left( 1+ 2 c_V\, \frac{h}{v} +   \cdots \right)   \\[0.15cm]
& - \frac{v}{\sqrt{2}} \lambda^u_{ij} \; \big( \bar u_L^{(i)} , \bar d_{L}^{(i)} \big) \, \Sigma \,   \big(u^{(i)}_R, 0\big)^T
       \left( 1+ c_u\, \frac{h}{v} + \cdots \right)+h.c. \\[0.15cm]
& - \frac{v}{\sqrt{2}} \lambda^d_{ij} \; \big( \bar u_L^{(i)} , \bar d_{L}^{(i)} \big)  \, \Sigma \,   \big(0, d^{(i)}_R\big)^T
       \left( 1+ c_d\, \frac{h}{v} + \cdots \right)+h.c. \\[0.15cm]
& - \frac{v}{\sqrt{2}} \lambda^l_{ij} \; \big( \bar \nu_L^{(i)} , \bar l_{L}^{(i)} \big)  \, \Sigma \,   \big(0, l^{(i)}_R\big)^T
       \left( 1+ c_l\, \frac{h}{v}  + \cdots \right)+h.c. 
\end{split}
\end{equation}
where the dots stand for terms with two or more Higgs fields and an  implicit sum over flavor indices $i,j = 1,2,3$
has been understood.
After rotating to the fermion mass eigenbasis and by choosing 
the unitary gauge $\Sigma(x) = 1$, the sum of (\ref{eq:V}) and (\ref{eq:L2}) coincides with the first two lines of Eq.~(\ref{eq:chiralL}) with $c_W = c_Z = c_V$.

At the level of four derivatives, there are 6 independent bosonic operators~\footnote{The operator 
${O}_2= (v/m_W^2) \, {\rm Tr} \!\left[ (D_\mu \Sigma)^\dagger (D_\nu \Sigma)\right] \partial^\mu \partial^\nu h$ 
that appeared in Eq.~(B.85)~
of the first version of this paper is actually redundant and can be eliminated in terms of the operators 
$O'_{WW}$ and $O'_{WB}$. For example, the shift $c'_{WW} \to c'_{WW} + c_2$ and $c'_{WB} \to c'_{WB} + \tan^2 \!\theta_W\, c_2$ can be used to remove $c_2$.} 
which affect cubic vertices with one $h$ 
field:~\footnote{Another convenient basis, which can be  more easily compared to Eq.~(\ref{eq:chiralL}), is
one in which the first two operators of Eq.~(\ref{eq:L4}) are replaced by
\begin{equation}
W^a_{\mu\nu} \,\Tr\!\left[ \Sigma^\dagger \sigma^a i\overleftrightarrow D^\mu \Sigma \right]  \partial^\nu h \, , 
\qquad
B_{\mu\nu} \,\Tr\!\left[ \Sigma^\dagger  i\overleftrightarrow D^\mu \Sigma \sigma^3 \right]  \partial^\nu h\, .
\end{equation}
This is in fact the basis adopted in Ref.~\cite{Azatov:2012bz}.
 }
\begin{equation}
\label{eq:L4}
\begin{split}
{\cal L}^{(4)} = 
&  \, c_{WW}^\prime\, W_{\mu\nu}^a W^{\mu\nu\, a}\, \frac{h}{v} 
    + c^\prime_{WB}\, \Tr\!\left( \Sigma^\dagger\, W_{\mu\nu}^a \sigma^a\, \Sigma\, B_{\mu\nu} \sigma^3 \right) \frac{h}{v} 
     + c_{BB}^\prime\, B_{\mu\nu} B^{\mu\nu} \, \frac{h}{v} \\[0.1cm]
& + \frac{c_W^\prime}{m_W}\, D^\mu W^a_{\mu\nu} \, \Tr\!\left( \Sigma^\dagger \sigma^a i\overleftrightarrow D_\nu \Sigma \right) h
    -\frac{c_B^\prime}{m_W}\, \partial^\mu B_{\mu\nu} \, \Tr\!\left( \Sigma^\dagger i\overleftrightarrow D_\nu \Sigma \, \sigma^3  \right) h \\[0.1cm]
& + \frac{c_{gg}}{2}\, G^a_{\mu\nu} G^{a\,\mu\nu} \, \frac{h}{v} + \dots
\end{split}
\end{equation}
The dots stand for terms which have two or more $h$ fields or do not lead to cubic vertices,
see Refs.~\cite{Azatov:2012bz,Alonso:2012px} for the  complete list of bosonic operators in ${\cal L}^{(4)}$.
In the unitary gauge, Eq.~(\ref{eq:L4}) coincides with the last three lines of Eq.~(\ref{eq:chiralL}).
More specifically, the coefficients $c_{WW}$, $c_{ZZ}$, $c_{Z\gamma}$, $c_{\gamma\gamma}$ can be written as linear combinations of 
$c_{WW}^\prime$, $c_{BB}^\prime$, $c_{WB}^\prime$, 
\begin{equation}
\label{eq:relationsforchiralL}
\begin{split}
c_{WW} & = 2 \, c'_{WW} \\
c_{ZZ} & = 2 (\cos^2\!\theta_W \, c'_{WW} - 2 \sin \theta_W \cos \theta_W \, c'_{WB} + \sin^2\!\theta_W \, c'_{BB})  \\
c_{\gamma\gamma} & =    2 (\sin^2\!\theta_W \, c'_{WW} + 2 \sin \theta_W \cos \theta_W \, c'_{WB} + \cos^2\!\theta_W \, c'_{BB}) \\
c_{Z\gamma} & = 2 (\sin \theta_W \cos \theta_W\, c'_{WW}  + \cos 2 \theta_W \, c'_{WB} - \sin \theta_W \cos \theta_W\, c'_{BB})\, ,
\end{split}
\end{equation}
while $c_{W\partial W}$, $c_{Z\partial Z}$ can be expressed in terms of $c_W^\prime$, $c_B^\prime$:
\begin{equation}
\label{eq:relationsforchiralL2}
\begin{split}
c_{W\partial W} & = 4  c'_{W} \\
c_{Z\partial Z} & = 4 \, c'_{W} + 4 \tan\theta_W \, c'_{B}  \\
c_{Z\partial \gamma} & = 4 \tan\theta_W \, c'_{W} - 4 \, c'_{B} \, .
\end{split}
\end{equation}
Notice that Eqs.~(\ref{eq:relationsforchiralL}) and (\ref{eq:relationsforchiralL2}) are directly implied by
Eq.~(\ref{eq:VVhatff}), which follows from custodial 
invariance.  It is simple  to verify that the identities~(\ref{eq:identity1}) and (\ref{eq:identity2})
are satisfied by the 
couplings appearing on the left-hand sides of respectively Eq.~(\ref{eq:relationsforchiralL}) and~(\ref{eq:relationsforchiralL2}).

The above discussion shows explicitly that every operator in Eq.~(\ref{eq:chiralL}) can be dressed up with NG bosons and made manifestly invariant under 
local $SU(2)_L \times U(1)_Y$ transformations.~\footnote{Notice that $h$ is \textit{invariant} under $SU(2)_L\times SU(2)_R$ (hence $SU(2)_L\times U(1)_Y$) 
transformations. In the case in which $h$ belongs to an $SU(2)_L$ doublet $H$, this follows from the fact that $h$ parametrizes the norm of the doublet: 
$H^\dagger H = (v+h)^2/2$.}

The part of Eq.~(\ref{eq:chiralLforHiggs})  which does not depend on the Higgs field $h$ coincides with the non-linear chiral Lagrangian for 
$SU(2)_L\times U(1)_Y$~\cite{EWchiralL},
in the limit of exact custodial symmetry. This latter assumption can be relaxed by specifying the sources of explicit breaking of the custodial
symmetry, \textit{i.e.} its spurions, in terms of which one can construct additional operators formally invariant
under  $SU(2)_L\times U(1)_Y$ local transformations. For example, the list of operators that follows in the case in which custodial invariance
is broken by a field with the EW quantum numbers of hypercharge has been recently discussed in Ref.~\cite{Alonso:2012px}. Since the choice of quantum
numbers of the spurions is model-dependent (and in fact the strongest effects are expected to arise from the breaking due to the top
quark, rather than hypercharge),  we do not report here any particular list of operators, and prefer to refer to the existing literature for further
details.

\section{Relaxing the CP-even hypothesis}
\label{app:CPodd}

If one relaxes the hypothesis that $h$ is CP-even, there are six extra dimension-6 operators that need to be added to the 
effective Lagrangian~(\ref{eq:silh}):
\begin{equation}
\label{eq:silhCPodd}
\begin{split}
\Delta {\cal L}_{CP} =
\, &
\frac{i \tilde c_{HW} \, g}{m_W^2}\, (D^\mu H)^\dagger \sigma^i (D^\nu H) {\tilde W}_{\mu \nu}^i
+\frac{i\tilde c_{HB}\, g^\prime}{m_W^2}\, (D^\mu H)^\dagger (D^\nu H) {\tilde B}_{\mu \nu} 
\\[0.2cm]
&+\frac{\tilde c_\gamma\,  {g'}^2}{m_W^2}\, H^\dagger H B_{\mu\nu} {\tilde B}^{\mu\nu}
   +\frac{\tilde c_g \, g_S^2}{m_W^2}\, H^\dagger H G_{\mu\nu}^a {\tilde G}^{a\mu\nu}
\\[0.2cm]
\, &
+    \frac{\tilde c_{3W}\,  g^3}{m_W^2}\, \epsilon^{ijk} W_{\mu}^{i\, \nu} W_{\nu}^{j\, \rho} {\tilde W}_{\rho}^{k\, \mu} 
+ \frac{\tilde c_{3G}\,  g_S^3}{m_W^2}\, f^{abc} G_{\mu}^{a\, \nu} G_{\nu}^{b\, \rho} {\tilde G}_{\rho}^{c\, \mu} \, ,
\end{split}
\end{equation}
where the dual field strengths are defined as $\tilde F_{\mu\nu} = \frac{1}{2} \epsilon_{\mu\nu\rho\sigma} F^{\rho\sigma}$
for $F=W,B,G$ ($\epsilon$ is the totally antisymmetric tensor normalized to $\epsilon_{0123}=1$). 
Furthermore, the coefficients of the operators involving  fermions will be in general  complex numbers.

In the case of the effective chiral Lagrangian with $SU(2)_L \times U(1)_Y$ non-linearly realized, there are four additional
operators, to be added to those of Eq.~(\ref{eq:L4}), which can affect cubic vertices with one $h$ field:
\begin{equation}
\label{eq:CPoddchiralL}
\begin{split}
\Delta {\cal L}^{(4)}_{CP} = 
&  \, \tilde c_{WW}^\prime\, \tilde W_{\mu\nu}^a W^{\mu\nu\, a}\, \frac{h}{v} 
    + \tilde c^\prime_{WB}\, \Tr\!\left[ \Sigma^\dagger\, \tilde W_{\mu\nu}^a \sigma^a\, \Sigma\, B_{\mu\nu} \sigma^3 \right] \frac{h}{v} \\[0.1cm]
& + \tilde c_{BB}^\prime\, \tilde B_{\mu\nu} B^{\mu\nu} \, \frac{h}{v} + \frac{\tilde c_{gg}}{2}\, \tilde G^a_{\mu\nu} G^{a\mu\nu} \, \frac{h}{v} \, .
\end{split}
\end{equation}

In the unitary gauge, both Lagrangians  $\Delta {\cal L}_{CP}$ and $\Delta {\cal L}^{(4)}_{CP}$ are matched onto:
\begin{equation}
	\label{eq:CP}
\begin{split}
\Delta {\cal L}^{(4)}_{CP} = 
 & \left(  \tilde c_{WW}\,   W_{\mu\nu}^+ \tilde W^{-\mu\nu}  + \frac{\tilde c_{ZZ}}{2} \, Z_{\mu\nu}\tilde Z^{\mu\nu} + 
\tilde c_{Z\gamma} \, Z_{\mu\nu} \tilde \gamma^{\mu\nu}   + \frac{\tilde c_{\gamma\gamma}}{2}\, \gamma_{\mu\nu}\tilde \gamma^{\mu\nu} + \frac{\tilde c_{gg}}{2}\, G_{\mu\nu}^a\tilde G^{a\mu\nu} \right) \frac{h}{v}
 + \dots 
\end{split}
\end{equation}
When the EW symmetry is linearly realized, the coefficients of Eq.~(\ref{eq:CP}) are related to the Wilson coefficients of Eq.~(\ref{eq:silhCPodd}) through 
the same relations reported in Table~\ref{tab:coupvalues} with the simple exchange
$c_i \to \tilde c_i$ (and with $c_W=c_B=0$). In the non-linear case, 
$\tilde c_{WW}, \tilde c_{ZZ}, \tilde c_{\gamma\gamma}$ and $\tilde c_{Z\gamma}$ are given in terms of the Wilson coefficients of Eq.~(\ref{eq:CPoddchiralL}) by  
relations identical to the ones of Eq.~(\ref{eq:relationsforchiralL}) (with $c_i \to \tilde c_i$ and  $c_2=0$). Notice that the Bianchi identities ensure that $D_\mu \tilde V^{\mu\nu}=0$ and therefore there are no CP-odd analogues to the operators $O_{V\partial V}$.

In addition to the new operators of Eq. (\ref{eq:CPoddchiralL}), an imaginary value of the coupling $c_{W\partial W}$ also breaks the CP-invariance\footnote{We thank A.~Pomarol for helping us to understand the issue of this additional CP-odd coupling.}:
\begin{equation}
 i \, {\rm Im}(c_{W\partial W})\, (W^-_\nu D_\mu W^{+\mu\nu} - h.c.).
\end{equation}
In the non-unitary-gauge, this coupling originates from the operator
\begin{equation}
\epsilon^{abc}\, \Tr\! \left( \Sigma^\dagger \sigma^a \Sigma \sigma^3 \right)
D^\mu W_{\mu\nu}^b \, 
\, \Tr\!\left( \Sigma^\dagger \sigma^c i\overleftrightarrow D^\nu \Sigma \right) h.
\end{equation}
In the linear realization of the $SU(2)_L\times U(1)_Y$ gauge symmetry, this coupling cannot be obtained from a dimension-6 operator but it originates from the dimension-8 operator:
\begin{equation}
i \epsilon^{abc} \left( H^\dagger \sigma^a H\right) \left( H^\dagger \sigma^b \overleftrightarrow D_\nu H\right) \left( D_\mu W^{c\, \mu\nu} \right).
\end{equation}
Notice that this coupling violates both the CP-invariance and the custodial symmetry.
Therefore, as in the CP-even sector,  there is a one-to-one correspondence between the couplings $hVV$ obtained from dimension-6 operators built with an EW doublet and the couplings $hVV$ obtained at the order $p^4$ when the $SU(2)_L\times U(1)_Y$ symmetry is non-linearly realized, provided that the gauge fields couple to conserved currents.

Finally, it should also be noted that when the CP-invariance assumption in the Higgs sector is relaxed, the couplings~$c_{u,d,l}$ are allowed to take some 
complex values.

\section{Current bounds on  dimension-6 operators}
\label{app:EWfit}

In this Appendix we explain how we derived the bounds on the coefficients of the dimension-6 operators reported in Section~\ref{sec:bounds}.
For a given observable we construct a likelihood for the  coefficients $\bar c_i$ as follows:
\begin{equation}
L(\bar c_i) \propto \exp\left[  -(O_{SM}+\delta O(\bar c_i) - O_{exp})^2/(2 \, \Delta O^2_{exp})\right] \, ,
\end{equation}
where $O_{exp}\pm \Delta O_{exp}$ is the experimental value of the observable, $O_{SM}$ denotes its SM prediction and $\delta O(\bar c_i)$
is the correction due to the effective operators.
If several observables constrain the same coefficients $\bar c_i$, the global likelihood is constructed by multiplying those
of each observable. We include the theoretical uncertainty on the SM prediction by integrating over a nuisance parameter
whose distribution is appropriately chosen. We then quote the bound on a given coefficient by marginalizing over  the remaining ones.

Let us consider for example the bounds of Eqs.~(\ref{eq:fermbounds1}) and (\ref{eq:fermbounds2}).
To derive them we used the EW fit performed in Ref.~\cite{Baak:2012kk} by the GFitter collaboration, and constructed a likelihood for the various coefficients
by computing their contributions to the $Z$-pole observables. For the latter, we used the SM predictions and experimental inputs
reported in Table~1 of Ref.~\cite{Baak:2012kk}, treating  the uncertainties on the SM predictions as normally distributed.
We performed two separate fits: one on the coefficients of the operators involving the light quarks ($u,d,s$),
and one on those with charged leptons and heavy quarks ($c,b$). We thus neglected, for simplicity, the correlations between these two
sets of coefficients. 
The relevant observables  in the first fit are $\Gamma_{tot}$, $\sigma_{had}$ and $R_l$. They depend on the Wilson coefficients only through
the following linear combination:
\begin{equation}
\begin{split}
l = & \left(-\frac{1}{4} + \frac{1}{3}\sin^2\!\theta_W \! \right)(\bar c_{Hq1} - \bar c'_{Hq1}) 
+ \left(\frac{1}{4} - \frac{1}{6} \sin^2\!\theta_W \! \right) (\bar c_{Hq1} + \bar c'_{Hq1} + \bar c_{Hq2} + \bar c'_{Hq2}) \\[0.1cm]
& + \frac{1}{3} \sin^2\!\theta_W\,  \bar c_{Hu}   - \frac{1}{6}\sin^2\!\theta_W \left( \bar c_{Hd} + \bar c_{Hs} \right)\, ,
\end{split}
\end{equation}
which with 95\% probability is constrained to lie in the interval
\begin{equation}
\label{eq:boundonlincomb}
-0.63 \times 10^{-3} < l  <  1.2 \times 10^{-3} \, .
\end{equation}
Although there are no further observables at the $Z$-pole which can  resolve the degeneracy implied by this result, we
 thought it useful to report the limits that one obtains from Eq.~(\ref{eq:boundonlincomb}) by turning on one coefficient at the time. These
are the bounds reported in Eq.~(\ref{eq:fermbounds1}). 

The second fit, performed on the coefficients of the operators with leptons and heavy quarks, makes use of all the observables at the $Z$ pole
and counts 7 unknowns, specifically: $x_i = \{ (\bar c_{Hq2}-\bar c'_{Hq2}), \bar c_{Hc}, (\bar c_{Hq3}+ \bar c'_{Hq3}), \bar c_{Hb}, \bar c_{Hl}, (\bar c_{HL}+ \bar c'_{HL}), 
(\bar c_{HL} - \bar c'_{HL}) \}$. For simplicity we assume lepton universality, and thus take the coefficients $\bar c_{Hl}$, $\bar c_{HL}$, $\bar c'_{HL}$
to be the same for all the leptonic generations.
In terms of the above variables, the result of the fit is summarized by their central values $\bar x_i$, standard deviations $\sigma_i$ and 
by the correlation matrix $\rho_{ij}$:
\begin{equation}
\begin{split}
\bar c_{Hq2}-\bar c'_{Hq2} & =  (5.8  \pm 4.4 ) \times 10^{-3}\\
\bar c_{Hc} & = (5.9  \pm  8.5 ) \times 10^{-3} \\
\bar c_{Hq3}+ \bar c'_{Hq3} & = (-3.1  \pm 2.7 ) \times 10^{-3}  \\
\bar c_{Hb} & = (-3.5  \pm 1.3 ) \times 10^{-2}  \\
\bar c_{Hl} &= (1.6  \pm 5.4 ) \times 10^{-4} \\
\bar c_{HL}+ \bar c'_{HL} & = (7.6  \pm 5.2 ) \times 10^{-4}  \\
\bar c_{HL} - \bar c'_{HL} & = (5.5  \pm 15 )  \times 10^{-4} 
\end{split}
\end{equation}
\vspace{0.25cm}
\begin{equation}
\rho = \begin{pmatrix}
 1.0 & 0.74 & -0.037 & -0.072 & 0.24 & -0.057 & -0.14 \\
 0.74 & 1.0 & -0.078 & -0.085 & 0.11 & 0.15 & 0.030 \\
 -0.037 & -0.078 & 1.0 & 0.85 & -0.40 & -0.21 & 0.068 \\
 -0.072 & -0.085 & 0.85 & 1.0 & -0.40 & -0.33 & -0.0024 \\
 0.24 & 0.11 & -0.40 & -0.40 & 1.0 & 0.11 & 0.28 \\
 -0.057 & 0.15 & -0.21 & -0.33 & 0.11 & 1.0 & -0.35 \\
 -0.14 & 0.030 & 0.068 & -0.0024 & 0.28 & -0.35 & 1.0
\end{pmatrix}
\end{equation}
\\[0.2cm]
The limits of Eq.~(\ref{eq:fermbounds2}) have been obtained by making use of the above formulas and marginalizing over
all the coefficients except the one on which the bound is reported.

For the limits of Eqs.~(\ref{eq:eps1}) and (\ref{eq:eps3}) we have used the fit on $S$ and $T$ performed in Ref.~\cite{Baak:2012kk}, by marginalizing 
on one parameter to extract the bound on the other.

To derive Eq.~(\ref{eq:udEDM}) we have used the theoretical predictions of the EDM of the neutron and mercury given in 
Ref.~\cite{Pospelov:2005pr} in terms of the dipole moments of the quarks (see Eqs.~(2.12), (3.65) and (3.71) of Ref.~\cite{Pospelov:2005pr}), and the experimental 
results for these observables given respectively in  Ref.~\cite{Baker:2006ts} and Ref.~\cite{Griffith:2009zz}.  We included the theoretical errors
by assuming that they are uniformly distributed within the stated intervals.
Only two linear combinations of the coefficients $\bar c_i$ can be constrained in this way, since two are the observables at disposal:
\begin{equation}
\begin{split}
l_1 =& -\frac{2 m_d}{m_W^2} \left[  \text{Im}(\bar c_{dG})  + 1.3\, \text{Im}(\bar c_{dB} - \bar c_{dW} ) \right]
           - \frac{m_u}{m_W^2} \left[ \text{Im}(\bar c_{uG}) - 0.64\, \text{Im}(\bar c_{uB} + \bar c_{uW} )  \right] \\[0.3cm]
l_2 =& -\frac{2 m_u}{m_W^2} \, \text{Im}(\bar c_{uG}) + \frac{2 m_d}{m_W^2} \, \text{Im}(\bar c_{dG}) \, .
\end{split}
\end{equation}
Using $m_u = 2.3\,$MeV and $m_d =4.8\,$MeV  we obtain, with 95\% probability:
\begin{equation}
\begin{gathered}
-1.59 \times 10^{-12}\, \text{GeV}^{-1} < l_1 < 1.78  \times 10^{-12} \, \text{GeV}^{-1}   \\[0.2cm]
-1.82 \times 10^{-12} \, \text{GeV}^{-1}  < l_2 < 1.37  \times 10^{-12} \, \text{GeV}^{-1}   \, .
\end{gathered}
\end{equation}
From the above result, by turning on one coefficient at the time, one obtains the limits given in Eq.~(\ref{eq:udEDM}).
The bound on $\text{Im}(\bar c_{tG})$ of Eq.~(\ref{eq:topcEDM}) has been similarly derived from the neutron and mercury EDMs by following 
Ref.~\cite{Kamenik:2011dk} and making use of the formulas given there. 

The limits of Eq.~(\ref{eq:RElepdipole}) have been obtained from the experimental measurements of the electron~\cite{Hanneke:2010au} 
and muon~\cite{PDGreview} 
anomalous magnetic moments and their SM predictions (taken respectively from Ref.~\cite{Giudice:2012ms} and Refs.~\cite{PDGreview,Davier:2010nc}). 
In this case we have included the theoretical errors by assuming that they are normally distributed.
All the remaining bounds reported in Section~\ref{sec:bounds}, namely those of Eqs.~(\ref{eq:topEDM})-(\ref{eq:anomtopcoupl}) and Eq.~(\ref{eq:IMlepdipole}) 
have been obtained by simply  translating into our notation the results given in the references quoted in the text.



\begin{thebibliography}{99}

\bibitem{:2012gk}
  G.~Aad {\it et al.}  [ATLAS Collaboration],
  Phys.\ Lett.\ B {\bf 716} (2012) 1
  [arXiv:1207.7214 [hep-ex]].

\bibitem{:2012gu}
  S.~Chatrchyan {\it et al.}  [CMS Collaboration],
  Phys.\ Lett.\ B {\bf 716} (2012) 30
  [arXiv:1207.7235 [hep-ex]].

\bibitem{weinberg}
  S.~Weinberg,
  ``The quantum theory of fields.''
  Cambridge, UK: Univ. Pr. (1996) 489 p

\bibitem{SILH}
  G.~F.~Giudice, C.~Grojean, A.~Pomarol and R.~Rattazzi,
  JHEP {\bf 0706} (2007) 045
  [hep-ph/0703164].

\bibitem{compositeHiggs}
  D.~B.~Kaplan and H.~Georgi,
  Phys.\ Lett.\  B {\bf 136} (1984) 183.
S.~Dimopoulos and J.~Preskill,
  Nucl.\ Phys.\  B {\bf 199}, 206 (1982).
T.~Banks,
  Nucl.\ Phys.\  B {\bf 243}, 125 (1984).
D.~B.~Kaplan, H.~Georgi and S.~Dimopoulos,
  Phys.\ Lett.\  B {\bf 136}, 187 (1984).
H.~Georgi, D.~B.~Kaplan and P.~Galison,
  Phys.\ Lett.\  B {\bf 143}, 152 (1984).
H.~Georgi and D.~B.~Kaplan,
  Phys.\ Lett.\  B {\bf 145}, 216 (1984).
M.~J.~Dugan, H.~Georgi and D.~B.~Kaplan,
  Nucl.\ Phys.\  B {\bf 254}, 299 (1985).

\bibitem{Contino:2003ve}
  R.~Contino, Y.~Nomura and A.~Pomarol,
  Nucl.\ Phys.\ B {\bf 671} (2003) 148
  [hep-ph/0306259].

\bibitem{Agashe:2004rs}
  K.~Agashe, R.~Contino and A.~Pomarol,
  Nucl.\ Phys.\ B {\bf 719} (2005) 165
  [hep-ph/0412089].

\bibitem{hdecay}
  A.~Djouadi, J.~Kalinowski and M.~Spira,
  Comput.\ Phys.\ Commun.\  {\bf 108} (1998) 56
  [hep-ph/9704448].
A.~Djouadi, M.~M.~Muhlleitner and M.~Spira,
  Acta Phys.\ Polon.\ B {\bf 38} (2007) 635
  [hep-ph/0609292].

\bibitem{eHDECAYpaper}
R.~Contino, M.~Ghezzi, C.~Grojean, M.~M.~Muhlleitner  and M.~Spira,
``eHDECAY: an implementation of the Higgs effective Lagrangian into HDECAY'', work in progress.


\bibitem{Buchmuller:1985jz}
C.~J.~C.~Burges and H.~J.~Schnitzer,
  Nucl.\ Phys.\ B {\bf 228} (1983) 464;
C.~N.~Leung, S.~T.~Love and S.~Rao,
  Z.\ Phys.\ C {\bf 31} (1986) 433;
    W.~Buchmuller and D.~Wyler,
  Nucl.\ Phys.\ B {\bf 268} (1986) 621.

\bibitem{others}
R.~Rattazzi,
  Z.\ Phys.\ C {\bf 40} (1988) 605;
  B.~Grzadkowski, Z.~Hioki, K.~Ohkuma and J.~Wudka,
  Nucl.\ Phys.\ B {\bf 689} (2004) 108
  [hep-ph/0310159];
  P.~J.~Fox, Z.~Ligeti, M.~Papucci, G.~Perez and M.~D.~Schwartz,
  Phys.\ Rev.\ D {\bf 78} (2008) 054008
  [arXiv:0704.1482 [hep-ph]];
  J.~A.~Aguilar-Saavedra,
  Nucl.\ Phys.\ B {\bf 812} (2009) 181
  [arXiv:0811.3842 [hep-ph]];
  J.~A.~Aguilar-Saavedra,
  Nucl.\ Phys.\ B {\bf 821} (2009) 215
  [arXiv:0904.2387 [hep-ph]].

\bibitem{Grojean:2006nn}
  C.~Grojean, W.~Skiba and J.~Terning,
  Phys.\ Rev.\ D {\bf 73} (2006) 075008
  [hep-ph/0602154].
 
\bibitem{Grzadkowski:2010es}
  B.~Grzadkowski, M.~Iskrzynski, M.~Misiak and J.~Rosiek,
  JHEP {\bf 1010} (2010) 085
  [arXiv:1008.4884 [hep-ph]].

\bibitem{Barbieri:2004qk}
  R.~Barbieri, A.~Pomarol, R.~Rattazzi and A.~Strumia,
  Nucl.\ Phys.\ B {\bf 703} (2004) 127
  [hep-ph/0405040].

\bibitem{Barbieri:1992dk}
  R.~Barbieri,
  CERN-TH-6659-92.

\bibitem{PT}
M.~E.~Peskin and T.~Takeuchi,
Phys.\ Rev.\ D {\bf 46} (1992) 381.

\bibitem{Agashe:2009di}
  K.~Agashe and R.~Contino,
  Phys.\ Rev.\ D {\bf 80} (2009) 075016
  [arXiv:0906.1542 [hep-ph]].

\bibitem{Isidori:2013ez}
  G.~Isidori,
  arXiv:1302.0661 [hep-ph].

\bibitem{Isidori:2010kg}
  G.~Isidori, Y.~Nir and G.~Perez,
  Ann.\ Rev.\ Nucl.\ Part.\ Sci.\  {\bf 60} (2010) 355
  [arXiv:1002.0900 [hep-ph]].

\bibitem{Sikivie:1980hm}
  P.~Sikivie, L.~Susskind, M.~B.~Voloshin and V.~I.~Zakharov,
  Nucl.\ Phys.\ B {\bf 173} (1980) 189.

\bibitem{eps123}
G.~Altarelli and R.~Barbieri,
Phys.\ Lett.\ B {\bf 253}, 161 (1991);
G.~Altarelli, R.~Barbieri and S.~Jadach,
Nucl.\ Phys.\ B {\bf 369}, 3 (1992)
[Erratum-ibid.\ B {\bf 376}, 444 (1992)].

\bibitem{Baak:2012kk}
  M.~Baak, M.~Goebel, J.~Haller, A.~Hoecker, D.~Kennedy, R.~Kogler, K.~Moenig and M.~Schott {\it et al.},
  Eur.\ Phys.\ J.\ C {\bf 72} (2012) 2205
  [arXiv:1209.2716 [hep-ph]].

\bibitem{Redi:2011zi}
  M.~Redi and A.~Weiler,
  JHEP {\bf 1111} (2011) 108
  [arXiv:1106.6357 [hep-ph]].

\bibitem{Vignaroli:2012si}
  N.~Vignaroli,
  Phys.\ Rev.\ D {\bf 86} (2012) 115011
  [arXiv:1204.0478 [hep-ph]].

\bibitem{Pospelov:2005pr}
  M.~Pospelov and A.~Ritz,
  Annals Phys.\  {\bf 318} (2005) 119
  [hep-ph/0504231].

\bibitem{Paradisi:2009ey}
  P.~Paradisi and D.~M.~Straub,
  Phys.\ Lett.\ B {\bf 684} (2010) 147
  [arXiv:0906.4551 [hep-ph]].

\bibitem{Kamenik:2011dk}
  J.~F.~Kamenik, M.~Papucci and A.~Weiler,
  Phys.\ Rev.\ D {\bf 85} (2012) 071501
  [arXiv:1107.3143 [hep-ph]].

\bibitem{AguilarSaavedra:2011ct}
  J.~A.~Aguilar-Saavedra, N.~F.~Castro and A.~Onofre,
  Phys.\ Rev.\ D {\bf 83} (2011) 117301
  [arXiv:1105.0117 [hep-ph]].

\bibitem{PDGreview}
A.~Hoecker and W.J.~Marciano, PDG review on ``The Muon Anomalous Magnetic Moment'', in:
  J.~Beringer {\it et al.}  [Particle Data Group Collaboration],
  Phys.\ Rev.\ D {\bf 86}, 010001 (2012).

\bibitem{Davier:2010nc}
  M.~Davier, A.~Hoecker, B.~Malaescu and Z.~Zhang,
  Eur.\ Phys.\ J.\ C {\bf 71} (2011) 1515
   [Erratum-ibid.\ C {\bf 72} (2012) 1874]
  [arXiv:1010.4180 [hep-ph]].

\bibitem{Hanneke:2010au}
  D.~Hanneke, S.~F.~Hoogerheide and G.~Gabrielse,
  arXiv:1009.4831 [physics.atom-ph].

\bibitem{Giudice:2012ms}
  G.~F.~Giudice, P.~Paradisi and M.~Passera,
  JHEP {\bf 1211} (2012) 113
  [arXiv:1208.6583 [hep-ph]].

\bibitem{Bennett:2008dy}
  G.~W.~Bennett {\it et al.}  [Muon (g-2) Collaboration],
  Phys.\ Rev.\ D {\bf 80} (2009) 052008
  [arXiv:0811.1207 [hep-ex]].

\bibitem{Hudson:2011zz}
  J.~J.~Hudson, D.~M.~Kara, I.~J.~Smallman, B.~E.~Sauer, M.~R.~Tarbutt and E.~A.~Hinds,
  Nature {\bf 473} (2011) 493.
  D.~M.~Kara, I.~J.~Smallman, J.~J.~Hudson, B.~E.~Sauer, M.~R.~Tarbutt and E.~A.~Hinds,
  New J.\ Phys.\  {\bf 14} (2012) 103051
  [arXiv:1208.4507 [physics.atom-ph]].

\bibitem{princeton}
R. Rattazzi, talk at the workshop \textit{Physics at LHC: from Experiment to Theory}, Princeton University, March 21-24 2007

\bibitem{Contino:2010mh}
  R.~Contino, C.~Grojean, M.~Moretti, F.~Piccinini and R.~Rattazzi,
  JHEP {\bf 1005} (2010) 089
  [arXiv:1002.1011 [hep-ph]].

\bibitem{Contino:2006qr}
  R.~Contino, L.~Da Rold and A.~Pomarol,
  Phys.\ Rev.\ D {\bf 75} (2007) 055014
  [hep-ph/0612048].

\bibitem{Choi:2002jk}
  S.~Y.~Choi, D.~J.~Miller, 2, M.~M.~Muhlleitner and P.~M.~Zerwas,
  Phys.\ Lett.\ B {\bf 553} (2003) 61
  [hep-ph/0210077].
  
\bibitem{DeRujula:2010ys}
  A.~De Rujula, J.~Lykken, M.~Pierini, C.~Rogan and M.~Spiropulu,
  Phys.\ Rev.\ D {\bf 82} (2010) 013003
  [arXiv:1001.5300 [hep-ph]].

\bibitem{Bolognesi:2012mm}
  S.~Bolognesi, Y.~Gao, A.~V.~Gritsan, K.~Melnikov, M.~Schulze, N.~V.~Tran and A.~Whitbeck,
  Phys.\ Rev.\ D {\bf 86} (2012) 095031
  [arXiv:1208.4018 [hep-ph]].

\bibitem{WP}
A.~Azatov, A.~Falkowski, C.~Grojean and E. Kuflik, work in progress.

\bibitem{Ellis:1975ap}
  J.~R.~Ellis, M.~K.~Gaillard and D.~V.~Nanopoulos,
  Nucl.\ Phys.\ B {\bf 106} (1976) 292;
  M.~A.~Shifman, A.~I.~Vainshtein, M.~B.~Voloshin and V.~I.~Zakharov,
  Sov.\ J.\ Nucl.\ Phys.\  {\bf 30} (1979) 711
   [Yad.\ Fiz.\  {\bf 30} (1979) 1368].

\bibitem{Kniehl:1995tn}
  B.~A.~Kniehl and M.~Spira,
  Z.\ Phys.\ C {\bf 69} (1995) 77
  [hep-ph/9505225].

\bibitem{Falkowski:2007hz}
  A.~Falkowski,
  Phys.\ Rev.\ D {\bf 77} (2008) 055018
  [arXiv:0711.0828 [hep-ph]].

\bibitem{Low:2010mr}
  I.~Low and A.~Vichi,
  Phys.\ Rev.\ D {\bf 84} (2011) 045019
  [arXiv:1010.2753 [hep-ph]].

\bibitem{Azatov:2011qy}
  A.~Azatov and J.~Galloway,
  Phys.\ Rev.\ D {\bf 85} (2012) 055013
  [arXiv:1110.5646 [hep-ph]].

\bibitem{Gillioz:2012se}
  M.~Gillioz, R.~Grober, C.~Grojean, M.~Muhlleitner and E.~Salvioni,
  JHEP {\bf 1210} (2012) 004
  [arXiv:1206.7120 [hep-ph]].

\bibitem{Redi:2012uj}
  M.~Redi,
  Eur.\ Phys.\ J.\ C {\bf 72} (2012) 2030
  [arXiv:1203.4220 [hep-ph]].

\bibitem{Barbieri:2012tu}
  R.~Barbieri, D.~Buttazzo, F.~Sala, D.~M.~Straub and A.~Tesi,
  arXiv:1211.5085 [hep-ph].

\bibitem{Degrande:2010kt}
  C.~Degrande, J.--M.~Gerard, C.~Grojean, F.~Maltoni and G.~Servant,
  JHEP {\bf 1103} (2011) 125
  [arXiv:1010.6304 [hep-ph]];
  H.~Hesari and M.~M.~Najafabadi,
  arXiv:1207.0339 [hep-ph];
  C.~Englert, A.~Freitas, M.~Spira and P.~M.~Zerwas,
  arXiv:1210.2570 [hep-ph].
  
\bibitem{Degrande:2012gr}
  C.~Degrande, J.~M.~Gerard, C.~Grojean, F.~Maltoni and G.~Servant,
  JHEP {\bf 1207} (2012) 036
  [arXiv:1205.1065 [hep-ph]].

\bibitem{CCWZ}
  S.~R.~Coleman, J.~Wess and B.~Zumino,
  Phys.\ Rev.\  {\bf 177} (1969) 2239;
  C.~G.~Callan, Jr., S.~R.~Coleman, J.~Wess and B.~Zumino,
  Phys.\ Rev.\  {\bf 177} (1969) 2247.

\bibitem{Burgess:1992gz}
  C.~P.~Burgess and D.~London,
  hep-ph/9203215.

\bibitem{Azatov:2012bz}
  A.~Azatov, R.~Contino and J.~Galloway,
  JHEP {\bf 1204} (2012) 127
  [arXiv:1202.3415 [hep-ph]].

\bibitem{Alonso:2012px}
  R.~Alonso, M.~B.~Gavela, L.~Merlo, S.~Rigolin and J.~Yepes,
  arXiv:1212.3305 [hep-ph].

\bibitem{WP2}
R.~Contino, C.~Grojean, D.~Pappadopulo, R.~Rattazzi, A.~Thamm, work in progress.

\bibitem{Passarino:2012cb}
  G.~Passarino,
  Nucl.\ Phys.\ B {\bf 868} (2013) 416
  [arXiv:1209.5538 [hep-ph]].

\bibitem{Weinberg:1978kz}
  S.~Weinberg,
  Physica A {\bf 96} (1979) 327.

\bibitem{Grojean:2013kd}
  C.~Grojean, E.~E.~Jenkins, A.~V.~Manohar and M.~Trott,
  arXiv:1301.2588 [hep-ph].

\bibitem{Elias-Miro:2013gya}
  J.~Elias-Miro, J.~R.~Espinosa, E.~Masso and A.~Pomarol,
  arXiv:1302.5661 [hep-ph].

\bibitem{Barbieri:2007bh}
  R.~Barbieri, B.~Bellazzini, V.~S.~Rychkov and A.~Varagnolo,
  Phys.\ Rev.\ D {\bf 76} (2007) 115008
  [arXiv:0706.0432 [hep-ph]].

\bibitem{Pomarol:2008bh}
  A.~Pomarol and J.~Serra,
  Phys.\ Rev.\ D {\bf 78} (2008) 074026
  [arXiv:0806.3247 [hep-ph]].

  \bibitem{Adler:1976zt}
 S.~L.~Adler, J.~C.~Collins and A.~Duncan,
 Phys.\ Rev.\ D {\bf 15} (1977) 1712.

\bibitem{Collins:1976yq}
 J.~C.~Collins, A.~Duncan and S.~D.~Joglekar,
 Phys.\ Rev.\ D {\bf 16} (1977) 438.

\bibitem{Nielsen:1977sy}
 N.~K.~Nielsen,
 Nucl.\ Phys.\ B {\bf 120} (1977) 212.

\bibitem{Farina:2012xp}
  M.~Farina, C.~Grojean, F.~Maltoni, E.~Salvioni and A.~Thamm,
  arXiv:1211.3736 [hep-ph].

\bibitem{Chetyrkin:1997un}
  K.~G.~Chetyrkin, B.~A.~Kniehl and M.~Steinhauser,
  Nucl.\ Phys.\ B {\bf 510} (1998) 61
  [hep-ph/9708255].

\bibitem{Kramer:1996iq}
  M.~Kramer, E.~Laenen and M.~Spira,
  Nucl.\ Phys.\ B {\bf 511} (1998) 523
  [hep-ph/9611272].

\bibitem{Schroder:2005hy}
  Y.~Schroder and M.~Steinhauser,
  JHEP {\bf 0601} (2006) 051
  [hep-ph/0512058].

\bibitem{Chetyrkin:2005ia}
  K.~G.~Chetyrkin, J.~H.~Kuhn and C.~Sturm,
  Nucl.\ Phys.\ B {\bf 744} (2006) 121
  [hep-ph/0512060].

\bibitem{nloggqcd}
  T.~Inami, T.~Kubota and Y.~Okada,
  Z.\ Phys.\ C {\bf 18} (1983) 69;
  A.~Djouadi, M.~Spira and P.~M.~Zerwas,
  Phys.\ Lett.\ B {\bf 264} (1991) 440;
  K.~G.~Chetyrkin, B.~A.~Kniehl and M.~Steinhauser,
  Phys.\ Rev.\ Lett.\  {\bf 79} (1997) 353
  [hep-ph/9705240].

\bibitem{Spira:1995rr}
  M.~Spira, A.~Djouadi, D.~Graudenz and P.~M.~Zerwas,
  Nucl.\ Phys.\ B {\bf 453} (1995) 17
  [hep-ph/9504378].

\bibitem{Baikov:2006ch}
  P.~A.~Baikov and K.~G.~Chetyrkin,
  Phys.\ Rev.\ Lett.\  {\bf 97} (2006) 061803
  [hep-ph/0604194].

\bibitem{Gunion:2002zf}
  J.~F.~Gunion and H.~E.~Haber,
  Phys.\ Rev.\ D {\bf 67} (2003) 075019
  [hep-ph/0207010].

\bibitem{Randall:2007as}
  L.~Randall,
  JHEP {\bf 0802} (2008) 084
  [arXiv:0711.4360 [hep-ph]].

\bibitem{Gunion:1984yn}
 J.~F.~Gunion and H.~E.~Haber,
 Nucl.\ Phys.\ B {\bf 272} (1986) 1
  [Erratum-ibid.\ B {\bf 402} (1993) 567].

\bibitem{Blum:2012ii}
  K.~Blum, R.~T.~D'Agnolo and J.~Fan,
  JHEP {\bf 1301} (2013) 057
  [arXiv:1206.5303 [hep-ph]].

  \bibitem{Azatov:2012qz}
  A.~Azatov and J.~Galloway,
  Int.\ J.\ Mod.\ Phys.\ A {\bf 28} (2013) 1330004
  [arXiv:1212.1380 [hep-ph]].

\bibitem{EWchiralL}
  T.~Appelquist and C.~W.~Bernard,
  Phys.\ Rev.\ D {\bf 22} (1980) 200;
  A.~C.~Longhitano,
  Phys.\ Rev.\ D {\bf 22} (1980) 1166;
  A.~C.~Longhitano,
  Nucl.\ Phys.\ B {\bf 188} (1981) 118.

\bibitem{Baker:2006ts}
  C.~A.~Baker, D.~D.~Doyle, P.~Geltenbort, K.~Green, M.~G.~D.~van der Grinten, P.~G.~Harris, P.~Iaydjiev and S.~N.~Ivanov {\it et al.},
  Phys.\ Rev.\ Lett.\  {\bf 97} (2006) 131801
  [hep-ex/0602020].

\bibitem{Griffith:2009zz}
  W.~C.~Griffith, M.~D.~Swallows, T.~H.~Loftus, M.~V.~Romalis, B.~R.~Heckel and E.~N.~Fortson,
  Phys.\ Rev.\ Lett.\  {\bf 102} (2009) 101601.

















\end{thebibliography}
\end{document}